\documentclass{aa}

\newcommand{\Pe}{\mathrm{Pe}}
\newcommand{\Pm}{\mathrm{Pm}}
\renewcommand{\Re}{\mathrm{Re}}
\newcommand{\Rm}{\mathrm{Rm}}
\renewcommand*\vec[1]{\ensuremath{\boldsymbol{#1}}}
\newcommand{\nel}{n_\mathrm{e}}
\newcommand{\mH}{m_\mathrm{H}}
\newcommand{\rin}{r_\mathrm{in}}
\newcommand{\rout}{r_\mathrm{out}}
\newcommand{\vphi}{\varphi}
\newcommand{\tdyn}{t_\mathrm{dyn}}
\newcommand{\vv}{\varv}
\newcommand{\vvv}{\vec{\varv}}
\newcommand{\qB}{\vec{q}_\mathrm{B}}
\newcommand{\kB}{k_\mathrm{B}}
\newcommand{\fb}{f_\mathrm{b}}

\newcommand{\hatb}{\hat{\vec{b}}}
\newcommand{\vnabla}{\vec{\nabla}}
\newcommand{\cV}{{\mathcal{V}}}
\newcommand{\cVn}[1]{{\mathcal{V}^{#1}}}
\newcommand*\abs[1]{\left|#1\right|}
\newcommand*\volave[1]{\langle#1\rangle_{\cV}}
\newcommand*\sphave[1]{\langle#1\rangle_{\Omega}}

\newcommand{\vir}{\mathrm{vir}}
\newcommand{\kappaeff}{\kappa_\mathrm{eff}}
\newcommand{\rms}{\mathrm{rms}}

\newcommand{\dd}{\mathrm{d}}
\newcommand{\zstroke}{%
  \text{\ooalign{\hidewidth -\kern-.3em-\hidewidth\cr$z$\cr}}%
}
\usepackage{natbib}
\usepackage{amsmath}
\usepackage{float}
\bibpunct{(}{)}{;}{a}{}{,}

\defcitealias{perrone2d}{pl22a}
\defcitealias{perrone3d}{pl22b}

\usepackage{graphicx}
\usepackage{txfonts}
\usepackage[breaklinks, colorlinks, citecolor=blue, urlcolor = blue]{hyperref}
\begin{document} 

   \title{Non-linear saturation and energy transport in global simulations of magneto-thermal turbulence in the stratified intracluster medium}

   \author{J. M. Kempf
          \inst{1}
          \and
          F. Rincon
          \inst{1}
          }

   \institute{$^1$Institut de Recherche en Astrophysique et Planétologie (IRAP), Université de Toulouse, CNRS, UPS, Toulouse, France\\
              \email{jean.kempf@irap.omp.eu}
             }

   \date{\today}

 
  \abstract
  {
  The magneto-thermal instability (MTI) is one of many possible drivers of stratified turbulence
  in the intracluster medium (ICM) outskirts of galaxy clusters,
  where the background temperature gradient is most likely aligned with the gravity.
  This instability occurs because of the fast anisotropic conduction of heat
  along magnetic field lines;
  but to what extent it impacts the ICM dynamics, energetics
  and overall equilibrium is still a matter of debate.
  }
  {
  This work aims at understanding MTI
  turbulence in an
  astrophysically stratified ICM atmosphere,
  its underlying saturation mechanism,
  and its ability to carry energy and to provide non-thermal pressure support.
  }
  {
  We perform a series of 2D and 3D numerical simulations of the MTI
  in global spherical models of stratified ICM,
  thanks to the finite-volume Godunov-type code IDEFIX,
  using Braginskii-magnetohydrodynamics. 
  We use well-controlled volume-, shell-averaged and spectral diagnostics
  to study the saturation mechanism of the MTI, and its radial transport energy budget.
  }
  {
  The MTI is found to saturate through a dominant balance between injection and dissipation
  of available potential energy,
  which amounts to marginalising the Braginskii heat flux
  but not the background temperature gradient itself.
  Accordingly, the strength and injection length of MTI-driven turbulence
  exhibit clear dependencies on the thermal diffusivity.
  With realistic Spitzer conductivity, the MTI drives cluster-size motions
  with Mach numbers up to $\mathcal{M} \sim 0.3$,
  even in presence of strong stable entropy stratification.
  We show that such mildly compressible flows can provide about $\sim 15\%$ of non-thermal pressure support
  in the outermost ICM regions close to the cluster accretion shock, and that the convective transport itself
  is much less efficient (a few percents only)
  than conduction at radially transporting energy.
  Finally, we show that the MTI saturation can be described 
  by a diffusive mixing-length theory,
  shedding light on the diffusive buoyant,
  rather than adiabatic convective, nature of the instability.
  }
  {
  The MTI seems relevant to both the dynamics and energetics of the ICM,
  through radially biased magnetic fields that enhance the background Braginskii heat flux.
  Further work including externally forced turbulence, mimicking accretion-induced turbulence for instance,
  is needed to assess its overall relative importance in comparison to other drivers of ICM turbulence. 
  }


   \titlerunning{Non-linear saturation and energy transport of MTI turbulence in the stratified ICM}
   \maketitle
%

\section{Introduction}
\label{sec:intro}

  The hot ($\kB T \sim 5 \ \mathrm{keV}$) and diffuse ($\rho \sim 10^{-27} \ \mathrm{gm/cm^3}$)
  intracluster medium (ICM) plasma
  is susceptible to the magneto-thermal instability \citep[MTI;][]{balbus00,balbus01,parrish05}
  in the periphery of galaxy clusters.
  This instability originates from the combination of
  features conducive to ICM outskirts:
  the dilute nature of its constitutive plasma
  and the inward alignment of the background temperature gradient with gravity.

  Despite the seemingly weak magnetic field filling galaxy clusters
  \citep[$B\sim 1-10 \ \mathrm{\mu G}$;][]{govoni04,ferrari08},
  ICM plasma is strongly magnetised:
  the Coulomb mean-free path of charged particles is fourteen orders of magnitude larger
  than their gyration radii around the local magnetic field. 
  This peculiarity makes the macroscopic transport
  of heat and momentum
  anisotropic with respect to the magnetic field lines \citep{braginskii65},
  which in turn strongly alters the buoyant response to perturbations
  of a stratified plasma layer
  threaded by weak magnetic fields.
  \citet{balbus00,balbus01,quataert08} showed
  that the buoyant stability of such an atmosphere
  is no longer controlled by the sign of the entropy gradient
  \citep[$ N^2 > 0$, with $N$ the Brunt-Väisälä frequency, as given by Schwarzschild's criterion;][]{schwarzschild06}
  but by that of the temperature gradient.
  In particular, despite the stable entropy stratification
  of galaxy clusters \citep{cavagnolo09},
  the ICM is either unstable to the MTI when the background temperature gradient is negative
  with respect to the radial direction \citep{balbus00,balbus01}
  or to its sister instability the heat-flux-driven buoyancy instability \citep[HBI;][]{quataert08}
  in the opposite case,
  both in the limit of fast parallel conduction
  and weak magnetic field.
  The MTI is therefore relevant to the periphery of galaxy clusters;
  while the HBI is more likely to develop in the cool-core of relaxed clusters,
  which exhibits increasing temperature with radius \citep{cavagnolo09}.
  Negative radial temperature gradients in galaxy clusters are actually ubiquitous,
  as observed in X-rays for example \citep{pratt07,leccardi08,ghirardini19},
  reflecting the hierarchical history of dynamical structure formation
  through gravitational accretion
  \citep{peterson06}.
  As shown by \citet{mccourt13}, temperature profiles observed
  in the Perseus cluster \citep{simionescu11},
  or in a larger sample of intermediate redshift clusters \citep{leccardi08},
  are a natural consequence of the cluster dynamical accretion history,
  modestly affected by the thermal conduction.
  The gravitational assembly process appears to be the main driver
  of the temperature distribution in galaxy clusters.

  Additionally, ICM turbulent fluid motions induced by such diffusive magneto-buoyant instabilities
  may induce a hydrostatic mass bias.
  The square of the Mach number $\mathcal{M}$,
  defined as the ratio between the fluid bulk velocity $\vv$ and the speed of sound
  $c_\mathrm{s} = \left(\gamma p/\rho\right)^{1/2}$
  with $\gamma$ the adiabatic index and $p$ the pressure,
  provides an estimate of the non-thermal turbulent pressure support
  found in galaxy clusters
  because it is proportional to the ratio of the dynamical to thermal pressures.
  \citet{parrish12b} argued that the MTI can reach Mach numbers
  as high as $\mathcal{M}\sim0.3$ and similar subsequent
  non-thermal to thermal pressure ratios
  at saturation in the extended outskirts of galaxy clusters.
  Such mildly compressible flows, if present in the outer ICM, would to some extent bias
  the X-ray and Sunyaev-Zeldovich (SZ) mass determination of galaxy clusters because these methods rely on
  the assumption of the ICM being at hydrostatic equilibrium
  \citep[HSE;][]{vazza18,eckert19,angelinelli20}.
  These possible mass biases are of significant cosmological importance
  since precise measurements of galaxy cluster mass distributions
  could help to alleviate, or at least to better understand,
  the current tension around the Hubble $H_0$
  and the $\sigma_8$ cosmological parameters
  \citep{allen11,pratt19}.
  
  Anisotropic thermal conduction, which is essential to the MTI,
  was first implemented in a large-scale structure formation simulation by \citet{ruszkowski10}.
  Since then, only \citet{ruszkowski11} specifically looked for any compelling signatures of the MTI in
  a simulation of structure formation through gravitational collapse;
  but they could not find any evidence of MTI-induced turbulence.
  This could either be due
  to the MTI being significantly inhibited
  by (viscous and/or perpendicular thermal) numerical diffusion,
  or to structure formation flows energetically outperforming the MTI.
  As a third option, \citet{ruszkowski11} suggested that
  any MTI flows, if present, would be hard to characterise given the very complex structure
  of the velocity field induced by gravitational collapse.
  A more detailed characterisation of MTI turbulence
  in high-resolution idealised global simulations, as first attempted by \citet{parrish08,parrish12b},
  is needed to bridge the gap between such models
  and large-scale structure formation simulations
  including anisotropic thermal conduction \citep{ruszkowski11},
  and to clarify the apparently conflicting conclusions reached by these different approaches.

  The issue of the MTI saturation level is all the more puzzling
  that two apparently different descriptions of the MTI turbulence at saturation
  have emerged during the past decade in works
  by \citet{parrish12b,mccourt13} and \citet{pl22a,pl22b}, respectively.
  The former claimed that MTI-driven turbulence can be described
  by a standard mixing-length theory involving the pressure scale-height
  $H_p = \left(\partial_r \log p\right)^{-1}$;
  while the latter invoked a locally balanced mechanism between
  the energy harvested from the background temperature gradient,
  and that dissipated by parallel thermal conduction.
  \citet{kempf23} suggested that which of these descriptions correctly depicts
  MTI-induced turbulence might depend on the typical injection length of the turbulence $\ell_i$,
  and more specifically, on how it compares with the typical scale-height $H_p$ of the cluster.
  A precise knowledge of the MTI turbulence strength and structure at saturation would help
  to better assess its detectability in large-scale simulations of structure formation \citep{ruszkowski11},
  or in the context of future X-ray observations of ICM outskirts dynamics \citep{kempf23}.
  The current study therefore also aims at clarifying this issue,
  thanks to global models of MTI in a stratified ICM atmosphere
  and phenomenological arguments.

  The outline of the paper is as follows.
  In Section \ref{sec:satMTI},
  we first recap the basic physics driving the instability.
  We then review the main descriptions proposed for the non-linear saturation of the MTI.
  In Section \ref{sec:model}, we present the numerical model,
  diagnostics,
  and set of global simulations we performed
  to study MTI-driven turbulence in presence of strong stable entropy stratification.
  These numerical simulations are then analysed in detail in Section \ref{sec:results},
  where we bring to light the saturation mechanism of the MTI in our models.
  We also document the dependency of the MTI structure and strength
  on the level of thermal conductivity at saturation.
  In Section \ref{sec:discussion},
  we introduce a diffusive mixing-length theory (MLT) of the MTI,
  and show how the two previously introduced descriptions of MTI-driven turbulence
  blend into a common framework.
  We then make use of this theory, and of the numerical results, to deduce
  the levels of convective flux and non-thermal turbulent pressure support
  that can be expected from MTI-driven turbulence in the outermost ICM regions.
  The main conclusions, caveats, and perspectives
  of our study are finally presented in Section \ref{sec:conclusion}.

\section{Phenomenology of the non-linear MTI}
\label{sec:satMTI}

  Before reviewing previously suggested descriptions of the MTI dynamics at saturation,
  we first recall the schematic linear mechanics of the instability.
  We picture a dynamically negligible horizontal magnetic field that threads a plasma
  layer with temperature gradient and gravity both pointing downwards.
  The initial magnetic field lines are therefore isothermal.
  When a fluid element is subsonically displaced from its initial equilibrium position,
  it remains thermally connected to its initial
  neighbours thanks to the dragging of the frozen-in magnetic field lines,
  in the limit of highly electrically conducting fluids.
  Provided that thermal conduction parallel to the magnetic field is fast
  enough, the perturbation is then isothermal. A rising blob will
  be hotter (and less dense) than the neighbouring fluid and
  keep rising through buoyancy, triggering the MTI.
  In the ICM, the latter can fully develop, with a maximal growth rate
  $\omega_T$ of about $0.5$ to $1 \ \mathrm{Gyr}^{-1}$,
  up to a state of saturated turbulence,
  where $\omega_T = \left(-g_0/H_T\right)^{1/2}$ is the MTI frequency,
  $g_0$ is the gravitational acceleration,
  and $H_T = \left(\partial_r \log T_0\right)^{-1}$ is the scale-height
  of the background temperature $T_0$.

  As discussed by \citet{kempf23},
  different physical mechanisms have been invoked
  to describe the turbulent saturation regime of the MTI.
  We chronologically review the two main schools of thought.
  On the one hand, \citet{parrish12b,mccourt13} 
  proposed to apply mixing-length arguments borrowed
  from the solar convection community
  \citep{vitense53,bohmvitense58,hurlburt84,chan89,canuto90,cattaneo91,abbett97,porter00}
  to predict both the intensity of temperature fluctuations
  \citep[and the subsequent level of convective heat flux;][]{mccourt13}
  and the kinetic energy \citep{parrish12b} induced by MTI-driven turbulence at saturation.
  The standard mixing-length theory (MLT), as applied to thermal convection,
  states that the temperature fluctuations
  induced by convective motions scale as \citep{vitense53}:
\begin{equation}
  \delta T \approx \left(\partial_r T_0 - \left.\partial_r T\right|_\mathrm{ad} \right)\ell_m,
\end{equation}
  where $\ell_m$ is the mixing length and
  $\left.\partial_r T\right|_\mathrm{ad}$ the adiabatic gradient,
  i.e. the temperature change of a fluid element
  due to adiabatic expansion (resp. compression) as it vertically moves in a stratified atmosphere.
  
  In this theory, the mixing length $\ell_m$ is taken to be the typical vertical length wandered
  by a blob of fluid before it loses its coherence and mixes with the background on account of diffusive effects.
  The incompleteness of the MLT lies in the choice of the free parameter $\ell_m$,
  which is typically taken to be a significant fraction $\alpha\sim 0.1$
  of the pressure scale-height $H_p$
  in stellar evolution models.
  Great efforts have been undertaken in the solar convection community
  to better constrain the mixing length thanks to numerical simulations
  \citep{chan89,abbett97,porter00}.
  In the MLT as applied to thermal convection, temperature perturbations scale as
  $\propto\left(\partial_r T_0 - \left.\partial_r T\right|_\mathrm{ad} \right)$
  (rather than simply $\propto\partial_r T_0$)
  because fluid elements are brought out of their equilibrium position
  adiabatically (rather than isothermally).
  In the case of the MTI however, initial perturbations
  are isothermal rather than adiabatic,
  on account of the fast parallel conduction along horizontal magnetic field lines:
\begin{equation}
  \frac{\delta T}{T_0} \approx \frac{\ell_m}{H_T}.
\label{eq:mlT}
\end{equation}
  A similar relation can be obtained for the velocity fluctuations,
  assuming that all the work performed by the buoyancy is converted into kinetic energy
  \citep{vitense53,abbett97,porter00}:
\begin{equation}
  \vv \approx \ell_m\omega_T,
\label{eq:mlv}
\end{equation}
  Therefore, $\vv$ must vary with $H_T$ (the precise power-law depending on how $\ell_m$ scales itself with $H_T$).
  This residual dependency was seen in the global numerical simulations of ICM from \citet{parrish12b},
  which certainly features cluster-size motions $\ell_m \sim H_p$.
  Absent a better physically motivated prescription for the mixing length $\ell_m$,
  these authors were not able to fully exploit Eqs. (\ref{eq:mlT})-(\ref{eq:mlv}).
  This limitation will be further explored, and overcome, in Section \ref{sec:dmlt}.

  On the other hand, \citet{pl22a,pl22b} argued that the MTI saturates
  once the energy locally injected by the instability
  (which critically depends on the thermal diffusion)
  into temperature fluctuations at a length-scale $\ell_i \ll H_T$,
  is dissipated by the parallel thermal diffusion itself.
  \citet{pl22a} showed that this balance can be formally rewritten:
\begin{equation}
  \frac{\nabla \delta T}{T_0} \approx H_T^{-1},
\label{eq:satMTI1}
\end{equation}
  and used to further deduce scaling laws for the MTI injection length and velocity fluctuations:
\begin{align}
  \label{eq:elli}
  &\ell_i \approx \left(\chi \omega_T\right)^{1/2}/N, \\
  \label{eq:vrms}
  &\vv \approx \left(\chi \omega_T^3\right)^{1/2}/N,
\end{align}
  the latter equation looking, at first glance, very different from Eq. (\ref{eq:mlv}),
  especially so if the mixing length $\ell_m$ is taken to be some fraction
  of the pressure scale-height $H_p$, as suggested by \citet{mccourt13}.
  We emphasise that the scenario described by \citet{pl22a}
  strongly relies on the energetics of the Boussinesq approximation,
  which looks however quite different from that of the compressible
  Braginskii-magnetohydrodynamics equations.
  Therefore, its leading energetic balance
  remains to be further theoretically understood and
  numerically tested in the context of MTI turbulence driven at global scales
  in a highly-stratified atmosphere representative of any astrophysical ICM.
  The resolution of this twofold problem is one of the main objectives of the present study.

  In both cases however, Eqs. (\ref{eq:mlT}) and (\ref{eq:satMTI1}) encapsulate the idea that
  steady-state turbulence is reached
  when temperature fluctuations rearrange so as to fight back the background temperature gradient,
  i.e. the source of free energy.
  Conversely, they are also the only dimensionally consistent way to translate this physical behaviour.
  The two differences between these theories are the privileged length scale
  and mode of heat transport invoked:
  the MLT from \citet{parrish12b,mccourt13} is based on the presumed convective aspect of the MTI,
  and has been deduced from global simulations of MTI-driven turbulence, in which $\ell_m\sim H_p$;
  while the theory of \citet{pl22a,pl22b} relies on the diffusive nature of the instability,
  and has been checked with Boussinesq simulations that inherently assume $\ell_i \ll H_T$.

  Accordingly, the first simulations of MTI-driven turbulence performed by \citet{parrish05,parrish07,parrish08}
  showed that the initial background temperature can be brought back to isothermality
  on cosmological time scales,
  at least when it is not continuously supplied in energy
  by boundary heat fluxes.
  In these simulations however, the respective contributions of convection and conduction in this
  isothermalisation process were not fully quantified.
  Under the homogeneous Boussinesq approximation used by \citet{pl22a,pl22b},
  no similar conclusions could be reached because the evolution of the background temperature profile
  was not self-consistently modelled.
  Consequently, ambiguous conclusions have been drawn regarding the ability of the MTI
  to efficiently carry heat through convective motions in the ICM. We will aim at clarifying this point
  in this paper too.
  In the same manner, which of Eqs. (\ref{eq:mlv})-(\ref{eq:vrms}) correctly describes the MTI velocity
  intensity at saturation may, critically or not, impact the level of non-thermal pressure support,
  and the subsequent estimates of hydrostatic mass bias,
  that can be expected from the turbulence induced by the instability.
  An objective of the current study is to provide further answers to this question.
  In the next section, we describe the numerical tools used
  to achieve these goals.

\section{Physical and numerical models}
\label{sec:model}

  We now present the set of Braginskii-magnetohydrodynamics (B-MHD) equations used in this study,
  and describe how they are normalised in
  the initial HSE model
  of a global stratified ICM common to all our simulations.
  Finally, we introduce the numerical methods used to solve these equations,
  and the energetics diagnostics used to analyse the simulations.

\begin{figure*}
\centering
\includegraphics[width=\hsize]{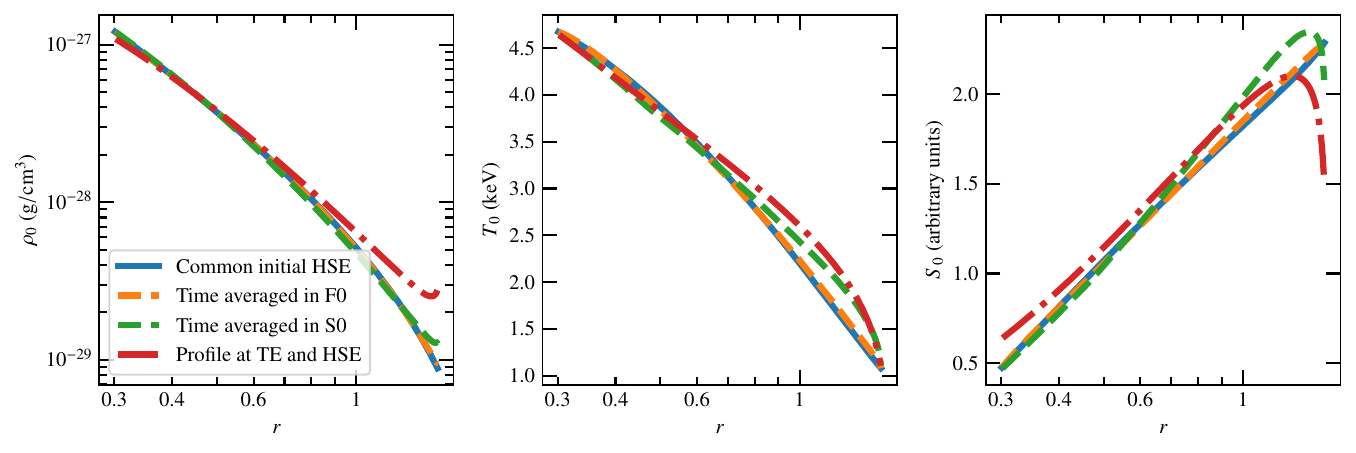}
\caption{
  In blue: initial stratified ICM atmosphere;
  density $\rho_0$ (in $\mathrm{g/cm^3}$, left), temperature $T_0$ (in $\mathrm{keV}$, center) and
  entropy $S_0$ (right) profiles.
  In yellow: same profiles but averaged over time for the 3D run F0 with $\kappaeff=0.075$.
  In green: same but for the 2D run S0 with a realistic Spitzer conduction (see discussion in Section \ref{sec:energyflux}).
  In red: for comparison only, an ICM atmosphere at both hydrostatic and thermal equilibria
  in presence of a hypothetical background heat flux with a Spitzer conductivity, Eq. (\ref{eq:spitzer}),
  subject to the same boundary conditions on the temperature. 
}
\label{fig:profiles}
\end{figure*}

\subsection{The Braginskii-magnetohydrodynamic equations}
\label{sec:bmhd}

  The compressible B-MHD equations are a system of non-linear partial differential equations
  describing the evolution of the density, the momentum and the energy
  of a perfectly ionised and dilute collisional plasma subject to anisotropic transport of
  heat and/or momentum (in a given spherical solution domain
  $\cV=\left(\vec{e_r}, \vec{e_\theta}, \vec{e_\varphi}\right)$ in the present study,
  see Section \ref{sec:numeric}).
  In conservative form, the set of equations is:
\begin{align}
\label{eq:densitycons}
  &\frac{\partial \rho}{\partial t}
  + \vnabla \cdot (\rho \vvv)
  = 0, \\
\label{eq:momentumcons}
  &\frac{\partial (\rho \vvv)}{\partial t}
  + \vnabla \cdot \left(\rho \vvv\vvv
  + \mathcal{P} \vec{I}
  - \frac{\vec{BB}}{\mu_0}
  - \vec{\Sigma}\right)
  =
  - \rho \vnabla\Phi, \\
\label{eq:Etotcons}
  &\frac{\partial E}{\partial t}
  + \vnabla \cdot \left( \left( E + \mathcal{P} \right) \vvv
  - \frac{\left(\vec{B} \cdot \vvv\right) \vec{B}}{\mu_0}
  - \frac{\eta \vec{B} \times \vec{j}}{\mu_0}
  - \vec{\Sigma}\cdot\vvv
  + \qB \right) \\ 
  &= 
  - \rho \vvv\cdot\vnabla\Phi, \nonumber \\
\label{eq:induction}
  &\frac{\partial \vec{B}}{\partial t}
  = \vnabla \times (\vvv \times \vec{B})
  + \eta \vec{\Delta B}.
\end{align}
  Here,
  $\rho$ is the mass density, $\vvv$ the fluid velocity (and $\vv$ its norm), $p$ the thermal pressure
  (i.e. the trace of the usual stress tensor).
  The vector $\vec{B}$ stands for the magnetic field (and the letter $B$ for its strength).
  The total pressure $\mathcal{P}=p+B^2/\left(2\mu_0\right)$
  is the sum of the hydrostatic and magnetic pressures
  (and $\mu_0$ is the vacuum permeability).
  The gravitational acceleration field $\vec{g}$, whose norm is $g_0$,
  equals the opposite gradient of the gravitational potential $\Phi$.
  The total energy volume density $E = E_K + E_M + E_I$ is the sum of the
  kinetic $E_K = \rho \vv^2/2$,
  magnetic $E_M = B^2/\left(2\mu_0\right)$ and
  internal $E_I = \mathcal{R}\rho T/\left(\gamma-1\right)$
  energy volume densities,
  where $T$ is the temperature of the gas
  and $\mathcal{R}$ the specific gas constant
  (i.e. the ratio between the Boltzmann constant and the mean molecular weight).
  We close the system with the usual equation of state for a perfect gas
  $p=\mathcal{R}\rho T$.
  The magnetic diffusivity is represented by the letter $\eta$ and
  the current density by the vector $\vec{j}=\vnabla\times\vec{B}/\mu_0$.

  The heat flux $\qB$ in a dilute, collisional, plasma
  is given by Braginskii's closure \citep{braginskii65}:
\begin{equation}
  \qB = -\kappa \left(\hatb\cdot\vnabla T\right) \hatb,
\label{eq:qbrag}
\end{equation}
  where $\kappa$ is the plasma thermal conductivity,
  and $\hatb$ is the unit vector in the direction of the magnetic field.
  In this study, we set the tensor $\vec{\Sigma}$ to be the isotropic viscous stress tensor:
\begin{equation}
  \vec{\Sigma} = \mu\left(\vnabla\vvv + \vnabla\vvv^T\right) -\frac{2}{3} \mu \left(\vnabla\cdot\vvv\right) \vec{\mathrm{I}},
\label{eq:sigma}
\end{equation}
  with $\mu$ the plasma dynamic viscosity.
  A complete description of the transport processes occurring in such plasmas
  would require the use of Braginskii viscosity \citep{parrish12a,kunz12}
  rather than the isotropic strain-rate tensor Eq. (\ref{eq:sigma}).
  Because the dynamics at the viscous scale
  is absent from both MTI theories reviewed in Section \ref{sec:satMTI},
  the dynamics at larger scales, in which we are mostly interested in this paper,
  should however be unaffected by this choice.
  Accordingly, we neglect the effects of kinetic micro-instabilities
  possibly triggered by Braginskii viscosity
  \citep[which is a disguised form of anisotropic pressure;][]{schekochihin05,schekochihin10},
  on the macroscopic heat transport
  \citep{riquelme16,komarov18,drake21},
  as they are not well described within B-MHD.
  In the dilute ICM,
  the plasma transport coefficients $\kappa$ and $\mu$
  are respectively set by the electrons and the protons \citep[][]{kunz12}.
  The plasma thermal conductivity is given by
  \citep{spitzer62,chandran04}:
\begin{equation}
   \kappa = 6.87\times10^{12} \left( \frac{k_\mathrm{B} T}{5 \ \mathrm{keV}} \right)^{5/2}
   \ \mathrm{g \ cm \ s^{-3} \ K^{-1}},
\label{eq:spitzer}
\end{equation}
  In the ICM, the Prandtl number,
  defined as the ratio between
  the plasma kinematic viscosity, $\nu=\mu/\left(\mH n_\mathrm{i}\right)$,
  and thermal diffusivity\footnote{In the present paper, $\kappa$ (resp. $\chi$) is a thermal conductivity (resp. diffusivity) in units of $\mathrm{g \ cm/s^3/K}$ (resp. $\mathrm{cm^2/s}$): their respective meanings are reversed with respect to \citet{kunz12,berlok21}. Our definition of the thermal diffusivity also differs from that of the previous authors by a factor $\nel/\left(\nel+n_\mathrm{i}\right)$, hence the different $\Pr$.}, $\chi=\kappa/\left(\nel\kB\right)$,
  is $\Pr\approx 0.01$ \citep{kunz12},
  where $\mH$ is the proton mass,
  and $\nel$, $n_\mathrm{i}$, the electron and ions number densities, respectively.

\subsection{Normalisation, hydrostatic equilibrium and parameters}
\label{sec:hse}

  In our simulations, lengths, times and densities are respectively normalised by the cluster
  virial radius $R_\vir$,
  the dynamical free-fall time $t_\mathrm{dyn}=\left(G M_\vir/R_\vir^3\right)^{-0.5}$
  and the third of the virial baryonic density
  $\rho_\vir/3=\fb M_\vir/(4\pi R_\vir^3)$,
  where $f_\mathrm{b}=0.17$ is the cosmic baryon fraction.
  In order to clarify expectations,
  the numerical values of these astrophysical scales are given in Tab. \ref{tab:norm},
  for the Perseus cluster \citep{simionescu11}.
  With such a choice,
  physical quantities other than lengths, times and densities are self-consistently normalised
  (through the B-MHD equations)
  by other typical cluster scales, defined and gathered in Tab. \ref{tab:norm}.
  For instance,
  the velocity is non-dimensionalised to the virial velocity
  $\vv_\vir=\left(G M_\vir/R_\vir\right)^{0.5}$
  and the temperature to twice the virial temperature $2\kB T_\vir=\mu m_\mathrm{H} \vv_\vir^2$.

  The 1D spherical hydrostatic ICM model of \citet{mccourt13} is used
  as a common initial condition for all the simulations of the paper\footnote{We have
  successfully reproduced this model,
  which has already been used and partially described in \citet{kempf23}.}.
  The corresponding density $\rho_0$ and temperature $T_0$ profiles
  are shown in blue in Fig. \ref{fig:profiles}.
  This spherically symmetric model
  efficiently reproduces some key thermodynamical features of the ICM
  as natural consequences of the cluster accretion history,
  like the non-zero temperature gradient
  despite the possible erasing effect of thermal conduction.
  More specifically, this model assumes fast relaxation of the ICM toward HSE 
  between successive accretion events of baryonic matter shells
  \citep[for a type III accretion history;][]{mcbride09}
  in a pre-settled and independent Navarro-Frenk-White \citep[NFW;][]{navarro97}
  gravitational potential of dark matter (we use a concentration parameter of 5),
  without any effect of the thermal conduction on the entropy evolution.
  Although thermal conduction can be accounted for in this 1D spherical model,
  the resulting temperature profiles are never at thermal equilibrium (TE)
  in presence of isotropic conductivity (or, equivalently, Braginskii with purely radial field lines),
  i.e. they do not satisfy
  $\partial_r(r^2 \kappa \partial_r T_0) = 0$ (for some given boundary conditions),
  because this condition is never imposed \textit{per se} in this model.
  For further reference though, an ICM atmosphere at both hydrostatic and thermal equilibria
  with a Spitzer conductivity, Eq. (\ref{eq:spitzer}), in a NFW potential well is shown
  in red in Fig. \ref{fig:profiles} too.
  This ICM atmosphere at both HSE and TE is qualitatively very similar to that of the
  \citet{mccourt13} model when a conductive heat flux is taken into account
  (see the red and yellow curves in their Fig. 7 and 9).
  In our simulations (which initially have $\partial_r(r^2 \kappa \partial_r T_0) \ne 0$),
  a modification of the background thermodynamic profile must be expected
  as soon as a non-negligible radial background heat flux operates.
  In Section \ref{sec:energyflux}, we will see that this adjustment manifests itself in simulations,
  after the opening of radial magnetic field lines by the onset of the MTI,
  through the development of a thermal wind,
  that remains even when a quasi-stationary state is reached,
  as previously noted by \citet{zingale02} in the case of an isotropic heat flux.
  In the linear phase though, no such dynamics is occurring and the initial background profiles are at rest.
  In agreement with observations, the atmosphere
  predicted by this model
  is stably stratified (with respect to the Schwarzschild criterion, $N^2=g_0/\left(\gamma H_S\right)>0$) in entropy
  $S_0 = \frac{\mathcal{R}}{\gamma-1}\log\left(T_0\rho_0^{1-\gamma}\right)$,
  with the ratio $N/\omega_T$ of the Brunt-Väisälä to the MTI frequencies of order unity,
  though slightly radii-dependent \citep{kempf23}.

  The MTI is excited through small initial white-noise velocity field perturbations,
  modelled by a gaussian random field with an amplitude of $10^{-4}$ for each component.
  The initial magnetic intensity is weak with $B_0=10^{-4}$;
  and the corresponding volume-averaged plasma beta parameter,
  defined as $\volave{\beta} = 2 \volave{p_0/B_0^2}$,
  is $\sim 10^{10}$.
  The associated magnetic configuration is
  a divergence-free, radially-modulated and purely azimuthal (i.e. horizontal) field,
  which is the magnetic geometry most unstable to the MTI:
\begin{equation}
  \vec{B} = B_0 \sin\left(\pi \frac{r-\rin}{\rout-\rin}\right) \vec{e_\vphi},
\label{eq:B0}
\end{equation}
  with $\rin=0.3$ and $\rout=1.5$
  (so that the magnetic field is initially zero at the inner and outer boundaries).
  In doing so, the background heat flux is initially shut down
  and the initial ICM atmosphere is at HSE.
  Then, we expect the background thermodynamic profile to stay still
  until an appreciable background heat flux appears due to the opening of the magnetic field lines by the MTI.

\begin{table}
\caption{
Set of typical cluster scales used to
non-dimensionalise the physical quantities of the simulations,
computed for the Perseus cluster \citep{simionescu11}.
}
\centering
\begin{tabular}{cccc}
\hline\hline
Physical quantity & Normalisation & Reference value \\
\hline
Length         & $R_\vir $ & $1.8$ Mpc \\
Time           & $t_\mathrm{dyn}$ & $1.4$ Gyr \\ 
Density        & $\rho_\vir/3$ & $1.04\times10^{-28} \ \mathrm{g/cm^{3}}$ \\
\hline
Velocity       & $\sqrt{G M_\vir/R_\vir}$ & $1260$ km/s \\
Thermal energy & $\mu \mH \vv_\vir^2$ & $10.3 \ \mathrm{keV}$ \\
Conductivity   & $\mathcal{R}\rho_\vir\vv_\vir R_\vir$  & $9.70\times10^{12} \ \mathrm{g \ cm/s^3/K}$ \\
Magnetic field & $\sqrt{\rho_\vir \mu_0}\vv_\vir$  & $45.6 \ \mathrm{\mu G}$  \\
\hline
\hline
\end{tabular}
\label{tab:norm}
\end{table}

\begin{table}
\caption{
List of 2D and 3D simulations, and their varying parameters.
}
\centering
\begin{tabular}{ccccc}
\hline\hline
Name & Resolution & $\kappa$-profile & $\kappa_0$ & $\kappaeff$ \\
\hline

3dSf4e0   & \small{256$\times$128$\times$512} & Spit. & 4.3 & 1 \\
2dSf4e0/S0   & 256 $\times$ 512 & Spit. & 4.3 & 1 \\
3dDf5e-2   & \small{256$\times$128$\times$512} & Cdiff. & 0.05 & 0.38 \\
2dDf5e-2   & 256 $\times$ 512 & Cdiff. & 0.05 & 0.38 \\
2dSf5e-1Mc   & 256 $\times$ 512 & Spit. & 0.5 & 0.12 \\
3dDf1e-2/F0   & \small{512$\times$256$\times$1024} & Cdiff. & 0.01 & 0.075 \\
2dDf1e-2   & 512 $\times$ 1024 & Cdiff. & 0.01 & 0.075 \\
2dSf6e-2   & 512 $\times$ 1024 & Spit. & 0.06 & 0.014 \\
2dSf3e-2Mc/S1   & 1024 $\times$ 2048 & Spit. & 0.03 & 0.0072 \\

\hline
\hline
\end{tabular}
\tablefoot{
All runs start with a purely azimuthal magnetic field $B_0=10^{-4}$,
and have $\Pr=0.01$, $\Pm=1$.
Spit (resp. Cdiff) stands for conduction profile $\kappa\propto T^{5/2}$ (resp. $\kappa\propto\rho$).
}
\label{tab:simu}
\end{table}

  The thermal diffusivity $\chi$
  is a key quantity of MTI-driven turbulence.
  Once the initial background density profile $\rho_0$ is chosen,
  the latter only depends on the thermal conductivity $\kappa$,
  and on the density fluctuations.
  To ensure that our conclusions are generic, two types of conduction profiles are explored in this paper:
  Spitzer (i.e. $\kappa \propto T^{\frac{5}{2}}$, noted Spit),
  for it is the most consistent with plasma physics,
  at least when kinetic effects arising in dilute plasma are not accounted for,
  and a type with constant diffusivity across radii (i.e. $\kappa \propto \rho$, noted Cdiff) for its simplicity.
  In both cases, we must specify the proportionality constant $\kappa_0$.
  For a Spitzer conduction profile, a natural choice appears:
  $\kappa_0 = 4.3$,
  which is the numerical value given by Eq. (\ref{eq:spitzer})
  for $T=2T_\vir$, further divided by $\kappa_\vir$.
  For the constant diffusivity profile, there is no such obvious option.
  For further comparison, we therefore introduce an additional dimensionless parameter $\kappaeff$
  which we define as the ratio between the volume-averaged conductivity in the solution domain
  (for any conduction type) and the volume average of the (full) Spitzer conductivity:
\begin{equation}
  \kappaeff = \frac{\volave{\kappa}}{\volave{4.3 T_0^\frac{5}{2}}}.
\end{equation}
  The Prandtl number $\Pr=\nu/\chi$,
  and its magnetic counterpart $\Pm=\nu/\eta$,
  are set to $0.01$ and $1$, respectively, in all simulations.
  The Brunt-Väisälä and the MTI frequencies, $N$ and $\omega_T$
  (two critical MTI timescales in the ICM),
  are enforced through the choice of the initial HSE,
  which is common to the main simulations of this study
  and reflects astrophysical ICM stratification regimes.
  We subsequently carry out an extensive parameter study,
  summarised in Tab. \ref{tab:simu},
  with respect to the profile and intensity of thermal conductivity only.

\subsection{Numerical methods, domains and boundary conditions}
\label{sec:numeric}
  We solve the system of Eqs. (\ref{eq:densitycons}-\ref{eq:induction}),
  normalised as described in Section \ref{sec:hse},
  using IDEFIX, a new Godunov-type finite-volume code for astrophysical flows \citep{lesur23}.
  At each cell interface, the in-and-out hyperbolic fluxes are computed through the approximate resolution 
  of the current Riemann problem with the HLLD solver
  \citep[which is a well-known compromise between numerical diffusion and stability;][]{miyoshi05}
  after the respective left and right states have been systematically reconstructed
  with a second-order scheme (associated to a slope limiter in order for the numerical scheme
  to mimic the total-variation-diminishing physical nature of the B-MHD equations).
  The Braginskii heat flux is computed with a finite-difference-like scheme \citep{parrish05}.
  We found that using a slope limiter on the anisotropic heat flux is mandatory
  to avoid episodic anti-diffusion \citep{sharma07}
  in a few runs, especially those in 2D in which the thermal conductivity has a Spitzer dependency.
  In the runs that have the suffix Mc in Tab. \ref{tab:simu},
  a monotonised central scheme ensures the monotonicity-preserving nature of the numerical operator accounting for anisotropic diffusion.
  Regarding the time-dependent part of the B-MHD equations,
  the hyperbolic terms are integrated thanks to a three-stage (third-order) Runge-Kutta algorithm,
  while the integration of parabolic operators is sped up using a Runge-Kutta-Legendre
  super time-stepping scheme \citep{meyer14, vaidya17}.
  The magnetic field is evolved thanks to a constrained transport algorithm \citep{fromang06},
  associated to the HLLD upwind constrained transport averaging scheme
  for the electromotive force \citep{londrillo04}.

  We now discuss the solution domain, alongside the boundary conditions and the numerical resolutions
  used in the simulations.
  We performed simulations in both two and three dimensions.
  In the former case, the B-MHD equations are solved in a curved domain $\cVn{2}$, which is a donut-like equatorial plane
  extending from 0.3 to 1.5 $R_\vir$
  (the usually expected position of the virial shock)
  in radius $r$,
  and from $0$ to $2\pi$ in azimuthal angle $\vphi$;
  while in three dimensions, $\cVn{3}$ is a spherical wedge
  extending over the same range of radii and azimuthal angles,
  and from $\theta=\pi/4$ to $3\pi/4$ in the polar direction.

  In two and three dimensions, the boundary conditions (BC) in the azimuthal direction are naturally periodic;
  while in three dimensions only, additional BCs are needed in the polar direction;
  we chose periodic BCs for all fields, making our 3D model semi-global.
  In both cases however, BCs are still needed in the radial direction.
  As mentioned in Section \ref{sec:intro},
  a numerical ICM atmosphere in which efficient MTI-driven turbulence is present
  can be quickly brought back to isothermality.
  Motivated by the non-marginal temperature gradients observed in astrophysical galaxy clusters,
  we indirectly impose Dirichlet BC on the temperature in the main simulations of this paper,
  by enforcing the pressure and the density to keep their initial HSE values in the ghost cells,
  so as to avoid this pathological behaviour.
  The consequence, on the temperature profile, of choosing inhomogeneous Dirichlet,
  rather than homogeneous Neumann, BCs is explored in detail in Section \ref{sec:energyflux}.
  Regarding the velocity, we use stress-free penetrative BC
  (i.e. homogeneous Neumann BC for both the normal and tangential components of the velocity field)
  at the inner edge of the domain, and a stress-free non-penetrative BC
  (i.e. homogeneous Dirichlet on the normal component)
  at the outer edge; such BCs seem more appropriate than penetrative BCs
  with respect to the physics of the accretion shock in astrophysical galaxy clusters.
  The impact of the velocity field BC on the overall dynamics of the simulation
  is quickly discussed in Section \ref{sec:energyflux}.
  At the inner and outer edges of the solution region,
  we impose that the tangential components of the magnetic field are zero
  (that is homogeneous Dirichlet BCs on the tangential components),
  while the radial component is automatically computed so as to satisfy the divergence-free nature of the field.
  Therefore, the magnetic field is purely radial (or null) at the inner and outer boundaries of the solution region,
  consistent with the initial magnetic geometry Eq. (\ref{eq:B0}).
  The uniform, non-adaptive resolutions
  of our 2D simulations range from $256\times512$ (i.e. $8.4$ kpc in the radial direction)
  to $1024\times2048$, to better resolve the runs with the lowest values of thermal diffusivity,
  as expected from the scaling of the MTI injection length, Eq. (\ref{eq:elli}).
  In 3D, the resolution of our simulations varies between $256\times128\times512$
  and $512\times256\times1024$.

\subsection{Energy budgets}
\label{sec:diag}
  We finally introduce the diagnostics used to characterise the turbulence generated by the MTI at saturation.
  The first quantities analysed are the averages of
  the Mach number $\volave{\mathcal{M}}$ and those of the
  available potential $\volave{E_A}$,
  kinetic $\volave{E_K}$, and magnetic $\volave{E_M}$, volume energy densities
  in the solution domain $\cV$.
  The available potential energy (APE) is defined as the amount of internal and gravitational potential energy
  which can be reversibly converted into kinetic energy;
  its budget will turn out to be particularly useful in Sec. \ref{sec:fidevolution}
  to understand how the MTI saturates.
  The use of this physical quantity is borrowed from environmental fluid dynamics,
  where it was first introduced by \citet{lorenz55},
  further developed by \citet{holliday81,andrews81,winters95},
  and is currently still intensively investigated by \citet{tailleux09,tailleux13,tailleux18}.
  Here, we simply define the APE volume density as:
\begin{equation}
  E_A = \frac{g_0^2}{2N^2} \frac{\delta \rho^2}{\sphave{\rho}},
\label{eq:quadape}
\end{equation}
  where we have used the local density fluctuation $\delta\rho = \rho - \sphave{\rho}$
  with respect to the spherically-averaged density $\sphave{\rho}$
  over solid angles $\Omega=\left(\theta,\vphi\right)$ at a given radius and time.
  From now on, spherical averages, denoted $\sphave{\cdot}$, of the thermodynamic quantities
  will formally replace the initial background profiles, denoted $\cdot_0$,
  in any qualitative discussion or quantitative diagnostic to come,
  because the background thermodynamic profiles are dynamically evolving through the course of the different simulations.
  To our knowledge, the notion of APE has never been directly mentioned in an astrophysical context,
  while for example
  a similar concept of available energy has been recently studied in the fusion plasma community
  \citep{helander17,helander20}.
  Nevertheless, we stress that
  a quadratic APE, as given by the right-hand side of Eq. (\ref{eq:quadape}),
  has often appeared (but has never been formally identified as such)
  in the context of astrophysical stratified fluid dynamics,
  like in studies of galaxy cluster turbulence \citep{pl22a,pl22b,mohapatra20,mohapatra21},
  of wave physics in stellar radiative zones \citep{prat14,ahuir21},
  or in planetary atmospheric dynamics \citep{dhouib24};
  precisely because the APE is exactly the part of the internal and gravitational potential energy
  that is modelled and accessible to a stratified fluid in the Boussinesq and anelastic approximations (through the buoyancy equation).
  Therefore, the leading energetic balance at the core of \citet{pl22a,pl22b} phenomenological picture
  of the MTI saturation mechanism is an APE budget in disguise.
  Additional details about the APE are presented in Appendix \ref{app:ape},
  here we just give its volume-averaged evolution equation in compressible B-MHD,
  assuming that no APE enters or leaves the domain
  and that the background profiles are constant in time:
\begin{align}
\label{eq:ape}
  \frac{\dd\volave{E_A}}{\dd t} =  &-\volave{\mathcal{A}} - \volave{\mathcal{C}} + \volave{\varepsilon_A} \\
  &+ \biggl \langle \frac{\delta T}{T} \vec{\Sigma}:\vnabla\vvv \biggr \rangle_\mathcal{V}
  + \biggl \langle \frac{\delta T}{T} \eta\,\mu_0\,j^2 \biggr \rangle_\mathcal{V}, \nonumber
\end{align}
  where $\delta p = p - \sphave{p}$ and $\delta T = T - \sphave{T}$ are the local pressure and temperature fluctuations.
  On the right-hand side of Eq. (\ref{eq:ape}), the first two terms quantify the amount
  of buoyant and compressible works,
  $\mathcal{A} = -g_0\delta\rho\delta\vv_r$ and $\mathcal{C} = -\vvv\cdot\vnabla\delta p$ respectively,
  that can reversibly convert APE into kinetic energy.
  The third term,
  $\varepsilon_A= \qB\cdot\vnabla\left(\frac{\delta T}{T}\right)$,
  is the conversion rate between
  the background potential energy (BPE, see Appendix \ref{app:ape} for an introduction) and the APE;
  the latter is created when $\varepsilon_A \geq 0$
  It can be further expanded into two contributions, $\varepsilon_i$ and $\varepsilon_\kappa$,
  by decomposing the temperature into its spherical average and fluctuating parts:
\begin{align}
\label{eq:epsi}
  &\varepsilon_i = -\kappa\,b_r \partial_r \sphave{T}\;\hatb\cdot\vnabla\left(\frac{\delta T}{T}\right), \\
\label{eq:epskappa}
  &\varepsilon_\kappa = -\kappa\,\left(\hatb\cdot\vnabla\delta T\right) \left(\hatb\cdot \vnabla\frac{\delta T}{T}\right).
\end{align}
  The nature of the fourth and fifth viscous and resistive terms on the right-hand side of Eq. (\ref{eq:ape}) are further discussed
  in Appendix \ref{app:ape}. They should be negligible on account of
  the regime of Prandtl and magnetic Prandtl numbers chosen in the simulations ($\Pr=0.01$, $\Pm=1$),
  and even more so for the resistive term in a realistic ICM since $\Pm\gg 1$ there \citep{schekochihin05}.

  Next, in order to better characterise energy transport within the simulation domain $\cV$,
  we introduce the evolution equation for the spherically-averaged total energy
  $E_I+E_K+E_M+E_G$, where $E_G=\rho\Phi$ is the gravitational potential energy:
\begin{align}
  \label{eq:radfluxes}
  &\frac{\partial}{\partial t} \sphave{E+E_G} \\
  &+ \frac{1}{r^2} \frac{\partial}{\partial r} \left( r^2 \left[
    \mathcal{Q}_r
  + \mathcal{H}_r
  + \mathcal{G}_r
  + \mathcal{K}_r
  + \mathcal{D}_r
  + \Pi_r
  \right] \right)
  = 0. \nonumber
\end{align}
  We refer the reader to Appendix \ref{app:energetics} for a description
  of the spherically-averaged kinetic $\mathcal{K}_r$, viscous $\mathcal{D}_r$, and Poynting $\Pi_r$ fluxes.
  We argue that, because of the very hot and gravitationally bounded nature of the ICM plasma, implying $E_M, E_K \ll E_I, \abs{E_G}$,
  these fluxes are sub-dominant with respect to
  the spherically-averaged radial Braginskii heat flux $\mathcal{Q}_r$,
  enthalpy flux $\mathcal{H}_r$ and gravitational flux $\mathcal{G}_r$:
\begin{align}
\label{eq:qr}
  &\mathcal{Q}_r = \sphave{-\kappa\left(\hatb\cdot\vnabla T\right) b_r}, \\
\label{eq:hr}
  &\mathcal{H}_r = \frac{\mathcal{R}\gamma}{\gamma-1}\sphave{\rho T}, \\
\label{eq:gr}
  &\mathcal{G}_r = \Phi\sphave{\rho\vv_r}.
\end{align}
  The radial Braginskii heat flux is the sum
  of the contributions from the average background temperature and from the temperature fluctuations: 
\begin{align}
\label{eq:qr0}
  &\mathcal{Q}_{r,0} = \sphave{-\kappa b_r^2}\partial_r\sphave{T}, \\
\label{eq:qr1}
  &\mathcal{Q}_{r,1} = \sphave{-\kappa\left(\hatb\cdot\vnabla \delta T\right) b_r}.
\end{align}
  The enthalpy and gravitational fluxes can similarly be subdivided into advective and convective contributions:
\begin{align}
\label{eq:hr0}
  &\mathcal{H}_{r,0} = \frac{\mathcal{R}\gamma}{\gamma-1}\sphave{\rho T}\sphave{\vv_r}, \\
\label{eq:hr1}
  &\mathcal{H}_{r,1} = \frac{\mathcal{R}\gamma}{\gamma-1}\left(\sphave{\rho}\sphave{\delta T \delta\vv_r}
                    +                                   \sphave{T}\sphave{\delta \rho \delta\vv_r}\right), \\
\label{eq:gr0}
  &\mathcal{G}_{r,0} = \Phi \sphave{\rho}\sphave{\vv_r}, \\
\label{eq:gr1}
  &\mathcal{G}_{r,1} = \Phi\sphave{\delta \rho \delta\vv_r}.
\end{align}
  The advective fluxes $\mathcal{H}_{r,0}, \ \mathcal{G}_{r,0}$ are related to a global radial mass transfer
  (which is zero in absence of any bulk motion $\sphave{\vv_r}$),
  whereas the convective fluxes
  $\mathcal{H}_{r,1} = \mathcal{H}_{r,\delta T} + \mathcal{H}_{r,\delta\rho}, \ \mathcal{G}_{r,1}$
  are due to a positive correlation between the turbulent fluctuations
  of temperature, density and radial velocity.
  Accordingly,
  we also monitor the volume-averaged level of the following cross-correlation function at saturation:
\begin{equation}
  \alpha\left(\delta\vv_r,\delta T\right) = \frac{\volave{\delta\vv_r\delta T}}{\sqrt{\volave{\delta\vv^2_r}\volave{\delta T^2}}}.
\end{equation}
  Finally, spectral diagnostics are used to assess both the convergence of the simulations
  and the spectral dynamics of MTI-driven turbulence.
  In 2D and 3D, we Fourier-analyse the azimuthal-$\vphi$ dependence,
  since spherical fields are naturally periodic in this direction,
  calling $m$ the associated angular degree.
  Additionally in 3D, we also do so for the polar-$\theta$ dependence
  ($l$ is the corresponding angular degree),
  as permitted by quasi-global $\theta$-periodic
  model used in this work.
  Therefore in 3D, the different fields are expanded as 2D plane waves
  $\propto\exp\left(\mathrm{i} l \theta + \mathrm{i} m \vphi \right)$ (and 1D plane waves $\propto\exp\left(\mathrm{i} m \vphi\right)$ in 2D runs).
  We stress that, in this work, $l$ is the angular degree of a plane wave and not the order of a spherical harmonic.
  We do not expand fields on a spectral basis in the radial direction,
  but we analyse the spectra of the radially-averaged angularly-expanded fields instead.
  Given the highly stochastic nature of turbulence, we systematically accumulate as much
  statistics as possible for the sake of diagnostic convergence.
  So, in the next section where simulation results are presented,
  all diagnostics are time-averaged
  (during the saturation phase of developed turbulence)
  when not plotted as a function of time.
  We caution however that such time averages are never made explicit in notations.
  All simulations are integrated during $200$ dynamical times $\tdyn$,
  knowing that the turbulent turnover time at saturation is $\sim 10 \tdyn$ in all simulations.
  
\section{Simulation results}
\label{sec:results}

  In this section, we present the results of the 2D and 3D simulations of MTI-driven turbulence
  whose parameters are summarised in Tab. \ref{tab:simu}.
  We first focus on the time evolution of the volume-averaged energy densities 
  and APE conversion rates of the fiducial 3D run F0.
  Then, we assess the structure and the intensity of the MTI flows, in all 2D and 3D simulations,
  using direct visualisation, global averages and spectral diagnostics.
  Finally, we analyse the energy flux budget in the quasi-stationary saturated state
  of MTI-driven turbulence in the 3D run F0.

\begin{table*}
\caption{
Key volume-averaged physical quantities at saturation for the 2D and 3D simulations
of MTI-driven turbulence listed in Tab. \ref{tab:simu}.
}
\centering
\begin{tabular}{cccccccccccccc}
\hline\hline
Run & $\Re$ & $\volave{\mathcal{M}}$ & $\volave{E_{K,r}}$ & $\volave{E_{K,h}}$ & $\volave{E_{M,r}}$ & $\volave{E_{M,h}}$ &
$\frac{\volave{\vv_r}}{\volave{\vv_{r,\rms}}}$ & $\volave{\varepsilon_i}$ & $-\volave{\varepsilon_\kappa}$ &
$m_i$ &
$\alpha\left(\delta\vv_r, \delta T\right)$ \\
\hline

3dSf4e0   & 100 & 0.14 & 0.0011 & 0.00058 & 4.1e-11 & 1.7e-11 & 0.61 & 0.0037 & 0.0028 & 3.1 & 0.59 \\
2dSf4e0/S0   & 67 & 0.17 & 0.0013 & 0.00091 & 9.9e-12 & 5.2e-12 & 0.59 & 0.0056 & 0.0047 & 4.1 & 0.54 \\
3dDf5e-2   & 170 & 0.13 & 0.001 & 0.0005 & 1.2e-09 & 5.7e-10 & 0.59 & 0.0017 & 0.0012 & 5.3 & 0.55 \\
2dDf5e-2   & 140 & 0.15 & 0.0011 & 0.00071 & 7.2e-11 & 5.8e-11 & 0.6 & 0.002 & 0.0016 & 5.9 & 0.43 \\
2dSf5e-1Mc   & 94 & 0.078 & 0.00021 & 0.00025 & 9.3e-10 & 1e-09 & 0.2 & 0.00075 & 0.00061 & 11.4 & 0.36 \\
3dDf1e-2/F0   & 111 & 0.05 & 0.00017 & 0.0001 & 1.3e-07 & 6.8e-08 & 0.35 & 0.00034 & 0.00024 & 15.4 & 0.5 \\
2dDf1e-2   & 119 & 0.065 & 0.00019 & 0.00026 & 3.4e-09 & 3.6e-09 & 0.27 & 0.0005 & 0.00041 & 12.8 & 0.29 \\
2dSf6e-2   & 126 & 0.032 & 2.9e-05 & 3.5e-05 & 1.1e-08 & 1.3e-08 & 0.065 & 8.4e-05 & 6.6e-05 & 30.1 & 0.28 \\
2dSf3e-2Mc/S1   & 123 & 0.023 & 1.6e-05 & 2e-05 & 2.1e-08 & 2.5e-08 & 0.048 & 4.4e-05 & 3.5e-05 & 43.4 & 0.25 \\

\hline
\hline
\end{tabular}
\tablefoot{
The angular degree $m_i$, representative of the injection scale, is defined in Eq. (\ref{eq:mi}).
In 2D, $E_{K,h} = E_{K,\varphi}$ while in 3D, $E_{K,h} = E_{K,\theta} + E_{K,\varphi}$,
and similarly for the magnetic energy $E_M$.
The Reynolds number is $\Re = \ell_i\volave{\vv_\rms}\volave{\rho}/\volave{\mu}$,
from which we further define the magnetic Reynolds number $\Rm=\Pm\Re=\Re$ and the Peclet number $\Pe=\Pr\Re=0.01\Re$.
}
\label{tab:2Dresult}
\end{table*}

\subsection{Evolution of the fiducial 3D MTI simulation}
\label{sec:fidevolution}
\begin{figure}
\centering
\includegraphics[width=0.9\hsize]{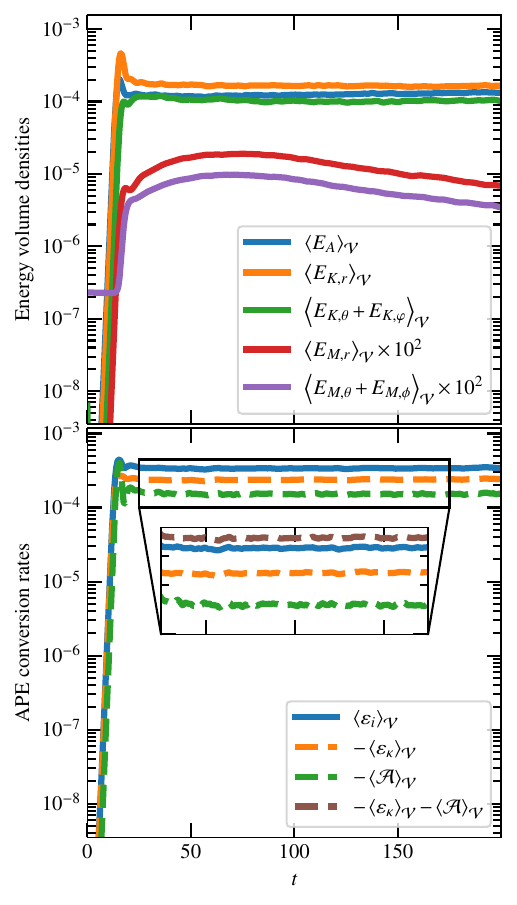}
\caption{
  Top: volume-averaged time evolution
  of the available potential energy density $\volave{E_A}$,
  and of the different contributions to the
  kinetic $\volave{E_K}$
  and magnetic $\volave{E_M}$ volume energy densities
  in the run F0.
  Bottom: volume-averaged time evolution of the
  APE injection rate $\volave{\varepsilon_i}$,
  APE thermal dissipation rate $-\volave{\varepsilon_\kappa}$,
  and the (opposite amount of) reversible buoyancy work $-\volave{\mathcal{A}}$
  during the 3D run F0.
  Inset: zoom-in on the highlighted region, in which the additional quantity
  $-\volave{\varepsilon_\kappa}-\volave{\mathcal{A}}$ is plotted.
  Dashed lines are indicative of negative quantities whose sign has been switched
  for the purpose of visualisation.
}
\label{fig:timeF0}
\end{figure}

  We first review the basic properties of the 3D MTI simulation F0,
  which has $\kappaeff = 0.075$.
  We choose it as our fiducial run because of its intermediate regime of thermal conductivity,
  and the simple shape of its diffusivity profile (that is constant across radii, like for all Cdiff runs).
  It will be later compared to the two 2D simulations, S0 and S1, with extreme values of thermal conductivity,
  and with a realistic Spitzer dependency,
  to emphasise the similar physics governing the dynamics of all these runs,
  no matter the regime, and dependency, of the thermal conductivity.
  The time evolution of the different component contributions to the average kinetic $\volave{E_K}$
  and magnetic $\volave{E_M}$ volume energy densities,
  and available potential energy (APE) $\volave{E_A}$,
  are shown on the top panel in Fig. \ref{fig:timeF0}.
\begin{figure*}
\centering
\includegraphics[width=\hsize]{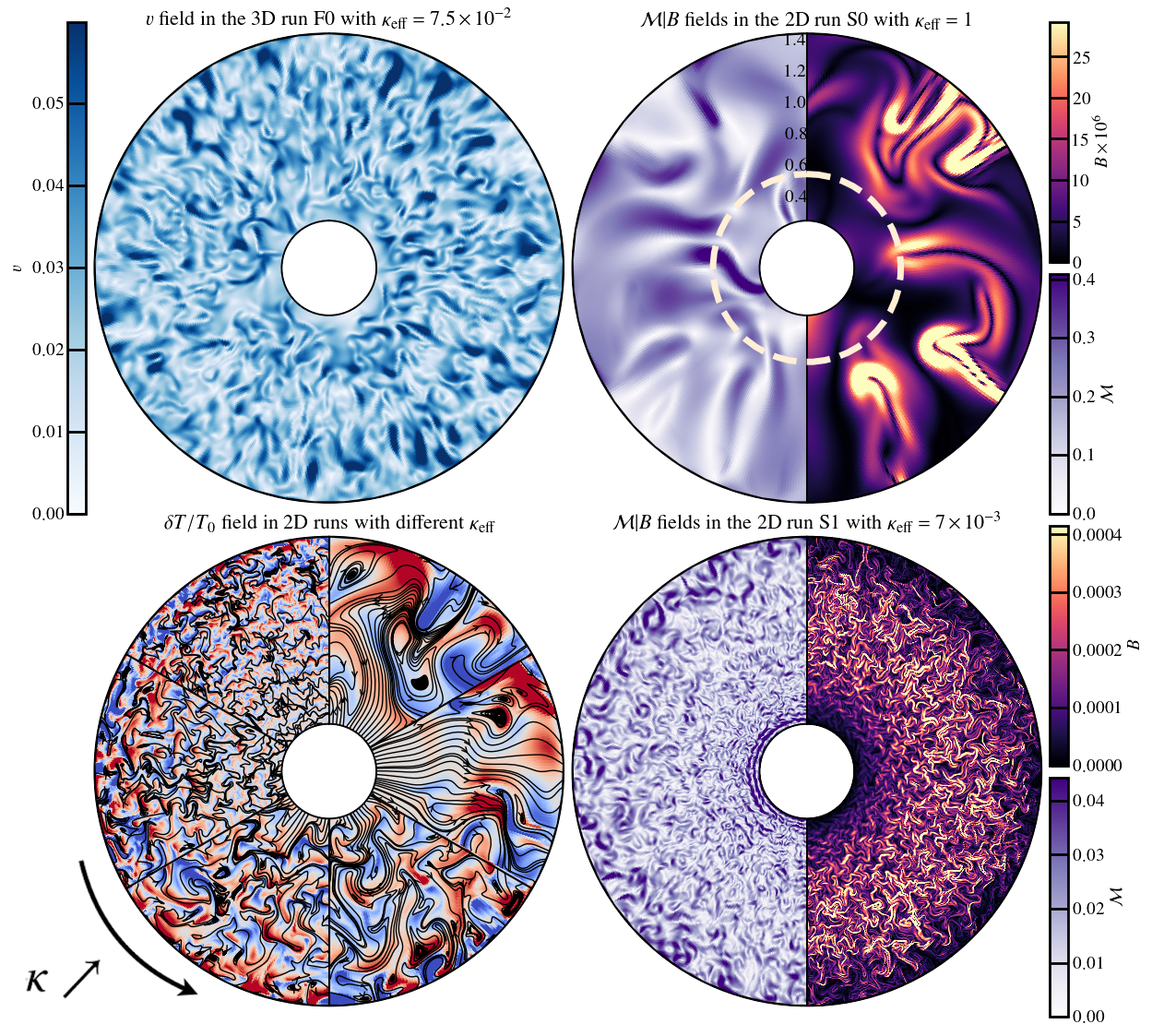}
\caption{
  Top left: velocity field in the equatorial plane $\left(r,\vphi\right)$ of the 3D run F0 at saturation.
  Top right: Mach number (left half) and magnetic strength (right half) snapshots of the 2D run S0 at saturation.
  Bottom right: same but for the 2D run S1.
  Bottom left: temperature fluctuation maps and magnetic field lines (in black).
  Different 2D runs are shown in each sixth of the pie,
  and are ranked
  (counter-clockwise, starting from the top left part)
  by increasing effective thermal conductivity $\kappaeff$,
  as in Tab. \ref{tab:simu}.
  In the latter plot, the color map is scaled independently for each sixth of the pie.
  Reddish blobs are hotter than the spherical average and blueish fluid elements are colder.
  The beige dashed line at $0.6R_\vir$ on the top right donut indicates the limit
  beyond which the ATHENA/X-IFU would collect less than $1\times10^6$ photons in the $0.2-12$ keV
  range during a $1$-Ms observation \citep[][see the discussion later in Sec. \ref{sec:implication}]{kempf23},
  as computed prior to the recent X-IFU design-to-cost exercise \citep{barret23}.
}
\label{fig:MTIfields}
\end{figure*}
  All these quantities quickly increase at the start of the simulation,
  as a result of MTI exponential growth.
  All along the simulation, the contribution of radial motions to the kinetic energy
  outweighs that of their horizontal counterparts,
  on account of the buoyant nature of the instability.
  Therefore, the MTI is able to drive energetic vertical motions,
  and thus despite the strong stable entropy stratification that tends to impede adiabatic fluid motions
  in the ICM \citep{mohapatra20,mohapatra21}.
  As highlighted in the first column in Tab. \ref{tab:2Dresult},
  the volume-averaged Reynolds number
  of the different simulations, including the fiducial 3D F0 run under consideration here,
  defined as $\Re=\ell_i\volave{\vv_\rms}\volave{\rho}/\volave{\mu}$
  (the injection length $\ell_i$ is quantified is the next subsection),
  is only moderate. Accordingly, the magnetic Reynolds number, defined as $\Rm = \Re\Pm$,
  is always lower than the critical threshold above which the MTI-driven turbulence can operate a
  dynamo\footnote{\citet{pl22b} found a critical $\Rm$ of about 35. Their definition however includes an additional $2\pi$ factor with respect to ours.} \citep{pl22b}.
  This is because of the $\Pm=1$ regime chosen here to avoid too constraining resolutions.
  Therefore, we do not expect significant feedback from the magnetic field
  through the (bulk) Lorentz force in these simulations,
  given the very high plasma beta of the initial state and the absence of dynamo, even in our 3D runs.
  Consequently, despite its initial exponential growth due to the MTI field-line stretching,
  the magnetic energy is several orders of magnitude smaller than the kinetic energy
  after the linear phase.
  The anisotropy between the radial and horizontal contributions to the magnetic energy
  is even stronger than for the kinetic energy.
  The slow, constant decrease of magnetic energy is due to magnetic field being advected out of the domain 
  (through the Poynting flux) by a thermal wind that is further described
  in Section \ref{sec:energyflux}.
  This dynamics however only modestly affects
  the turbulence here since all components are decaying at the same rate,
  and, in the B-MHD regime considered here,
  only the direction (and not the magnitude) of the magnetic field (through anisotropic heat flux) matters
  when the magnetic field stays dynamically negligible (which is the case here).

  To understand MTI saturation,
  we now focus on the time evolution of the different sink and source terms of APE,
  Eq. (\ref{eq:ape}).
  In the bottom panel of Fig. \ref{fig:timeF0}, we display the time evolution
  of the average conversion rates, $\volave{\varepsilon_i}$, $-\volave{\varepsilon_\kappa}$,
  along with the (opposite amount of) reversible buoyancy work $-\volave{\mathcal{A}}$.
  The first striking feature is the positive sign of $\volave{\varepsilon_i}$.
  \citet{middleton20,tailleux24} recently showed that, in oceanic physics,
  conversion from background potential energy (BPE) into APE can be associated to
  the growth of double-diffusive Ledoux-like instabilities.
  On the contrary, in incompressible Boussinesq single-component stratified fluid dynamics,
  APE can only be dissipated into BPE, and backward transferring from BPE to APE is prevented \citep{winters95,middleton20}.
  Therefore, such a positive conversion rate
  seems indicative of a thermodynamic process that harvests energy
  from the BPE
  and invests it into APE in the form of temperature and density fluctuations.
  The injected APE $\volave{\varepsilon_i}$ is then distributed between
  the amount of APE dissipated back to the BPE $\volave{\varepsilon_\kappa}$
  and the kinetic energy, through the reversible work of buoyancy $\volave{\mathcal{A}}$.
  Quantitatively speaking, the relation $\varepsilon_i+\varepsilon_\kappa+\mathcal{A}=0$ does not hold exactly
  (see the inset in Fig. \ref{fig:timeF0}),
  most certainly because of small APE boundary fluxes not accounted for in the current budget (see Appendix \ref{app:ape}).
  This suggests that the saturation mechanism
  proposed by \citet{pl22a,pl22b}
  holds for global stratified MTI-driven turbulence developing
  in an astrophysically stratified ICM atmosphere as well,
  as this is precisely the balance on which their theory relies
  ($\varepsilon_i\approx-\varepsilon_\kappa-\mathcal{A}$).
  We therefore expect the scaling laws for the MTI kinetic energy Eq. ($\ref{eq:vrms}$)
  and injection length Eq. ($\ref{eq:elli}$) to remain valid in global MTI-driven turbulence.
  As a final note, we highlight that
  $\varepsilon_i\approx-\varepsilon_\kappa$ is just a restatement of $\qB\approx 0$.
  Thus, we found that the MTI saturates by marginalising the Braginskii heat flux $\qB$,
  whereas \citet{parrish05,parrish07} argued
  that the MTI would saturate by shutting itself off
  through the marginalisation of the background temperature gradient $\partial_r \sphave{T}$.

\begin{figure}
\centering
\includegraphics[width=0.87\hsize]{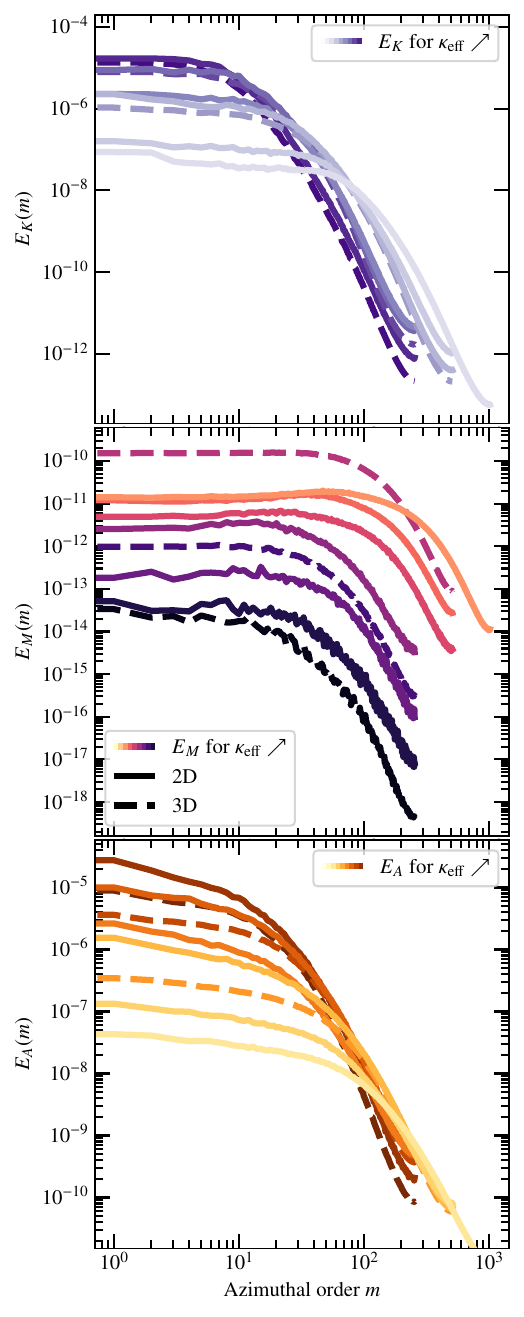}
\caption{
  Radially-averaged spectra plotted against the azimuthal order $m$,
  for all simulations in Tab. \ref{tab:simu},
  with darker colors corresponding to higher $\kappaeff$.
  Dotted lines are for 3D runs and regular lines for 2D runs.
  Top row: kinetic energy spectral density.
  Middle row: magnetic energy spectral density.
  Bottom row: APE spectral density.
}
\label{fig:spec2D}
\end{figure}

\subsection{Structure and strength of the flow}
\label{sec:flowstructure}

  In this section, we characterise the turbulent properties of MTI-induced turbulence,
  both in terms of strength and structure.
  On the top left donut in Fig. \ref{fig:MTIfields},
  we show the velocity field induced by the MTI at saturation in the 3D simulation F0,
  where buoyant plumes characteristic of the 3D MTI \citep{pl22b} are discernible.
  These turbulent eddies, which are preferentially elongated in the direction of gravity,
  reflect the anisotropy in the partition of kinetic energy between vertical and horizontal components,
  that was highlighted in the previous subsection.
  In all 2D and 3D simulation, the typical phenomenology of the MTI is respected:
  hot fluid elements rise while colder blobs sink,
  as quantified by the strong positive cross-correlation $\alpha\sim 0.5$
  between the radial velocity and temperature fluctuations
  (last entry in Tab. \ref{tab:2Dresult}).
  Additionally, the bottom left donut in Fig. \ref{fig:MTIfields} shows that
  strong gradients of temperature fluctuations, typical of the MTI,
  are developing perpendicular to the magnetic field lines
  (that are superimposed as black lines on slices of different 2D runs).

  As shown in Section \ref{sec:fidevolution},
  the MTI saturation mechanism of \citet{pl22a,pl22b} theory,
  $\varepsilon_i\approx-\varepsilon_\kappa\approx-\mathcal{A}$,
  essentially seems to hold
  in our global stratified models of MTI-driven turbulence.
  A key prediction of the theory is then the scaling of the injection length, Eq. (\ref{eq:elli}),
  and of the kinetic energy, Eq. (\ref{eq:vrms}), with the thermal diffusivity.
  Such a scaling could have important implications for the ICM
  given the large value of Spitzer conductivity in such plasma
  (if unsuppressed by kinetic instabilities).
  We therefore check both scaling laws,
  along with that for the APE,
  in global MTI-driven and stratified turbulence,
  on a volume-averaged basis for all runs in Tab. \ref{tab:simu}.
  In order to better assess the dependency of the flow strength and structure on the thermal diffusivity,
  we mostly rely on the 2D runs, which allow a broader parameter sweep than in 3D.
  We start with both 2D runs S0 and S1,
  which only differ in the intensity of their thermal conductivities:
  they have respectively the highest and the lowest ones.
  Despite sharing a similar phenomenology,
  MTI flows in S0 and S1 exhibit very different driving scales at saturation
  (which is easily seen by eye-comparison of both right donuts in Fig. \ref{fig:MTIfields}).
  For instance,
  cluster-size flows are present in S0 (top right donut),
  which corresponds to a realistic Spitzer conductivity case;
  while the typical turbulent length scale is much smaller in S1 (bottom right donut),
  and progressively increases 
  with the level of thermal conductivity (counter-clockwise bottom left donut).
  
  We can better quantify this diffusivity dependence using the spectral diagnostics
  defined in Section \ref{sec:diag}.
  In Fig. \ref{fig:spec2D}, we show the different radially-averaged spectral energy densities
  $E_K(m)$, $E_M(m)$ and $E_A(m)$
  for all runs in Tab. \ref{tab:simu}
  In 3D, an additional average across the polar angular degree $l$ is performed: $E_K(m) = \sum_l E_K(m,l) / n_l$, where $n_l$ is the number of polar modes.
  These plots illustrate the phenomenological picture of turbulence,
  in which energy is distributed over a range of scales,
  but most of it resides at large scales.
  The slopes are quite different from
  classical $k^{-5/3}$ (or $k^{-3}$ in 2D) Kolmogorov laws.
  This is expected for MTI turbulence in the ICM since it is neither homogeneous, nor isotropic.
  We further note that the kinetic and magnetic energies tend
  to reach smaller scales than the APE (i.e. the thermodynamic fluctuations),
  on account of the low thermal Prandtl number which makes the parallel thermal dissipation scale
  larger than the (isotropic) resistive and viscous scales.

\begin{figure}
\centering
\includegraphics[width=0.965\hsize]{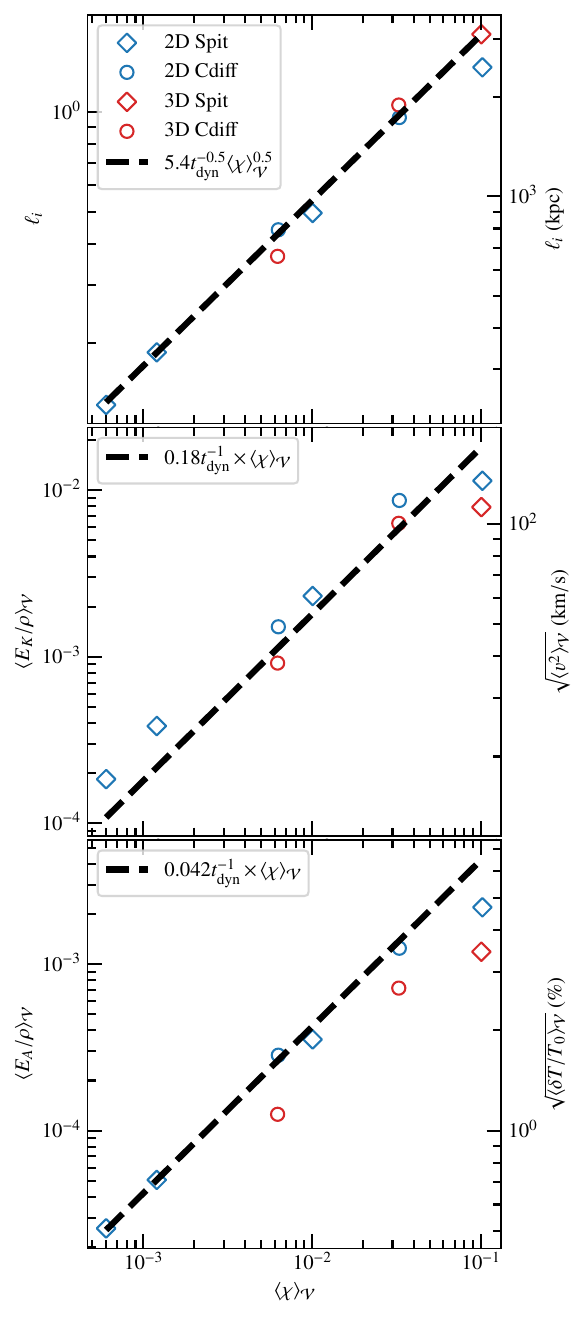}
\caption{
  MTI scaling laws.
  Volume average of the turbulent injection length (top),
  the velocity fluctuation strength (middle),
  and specific APE (bottom),
  as a function of the
  volume-averaged thermal diffusivity, for all simulations in Tab. \ref{tab:simu}.
  Diamonds (respectively, circles) correspond to runs with Spitzer
  (respectively, constant diffusivity) conduction profile;
  while the color blue (resp. red) stands for 2D (resp. 3D) runs.
}
\label{fig:scalings}
\end{figure}

  We are now in a position to constrain the MTI integral scale $\ell_i$,
  using the azimuthal angular degree $m_i$ (in both 2D and 3D) deduced from the radially-averaged volume kinetic energy spectral density $E_K(m)$ in 2D, $E_K(l,m)$ in 3D:
\begin{align}
\label{eq:mi}
  &m_i = \frac{\sum_m m E_K}{\sum_m E_K}, \ \mathrm{in \ 2D}, \\
\label{eq:mi3D}
  &m_i = \frac{\sum_m m \left(\sum_l E_K \right)}{\sum_m \left(\sum_l E_K\right)}, \ \mathrm{in \ 3D}.
\end{align}
  This quantity roughly matches the order $m$ at which the kinetic spectrum peaks.
  In the top panel of Fig. \ref{fig:scalings},
  we show the relation between the global horizontal injection length, defined as
  $\ell_i = \pi\left(\rin+\rout\right)/m_i$ (in both 2D and 3D)
  and the volume-averaged thermal diffusivity, for all simulations in Tab. \ref{tab:simu}.
  The scaling Eq. (\ref{eq:elli}) is very well verified
  across two decades of variation in thermal diffusivity.
  From an astrophysical viewpoint,
  this means that the MTI alone drives cluster-size flows in the ICM,
  with unsuppressed realistic Spitzer conductivity ($\kappaeff=1$),
  as illustrated in the top right donut in Fig. \ref{fig:MTIfields}.

  Next, we focus on the strength of the flow induced by MTI turbulence
  and compare it to the scaling Eq. (\ref{eq:vrms}).
  Again, this comparison is made in a global sense.
  On the middle panel in Fig. \ref{fig:scalings},
  we show the evolution of the volume-averaged squared velocity fluctuations
  as a function of the volume-averaged thermal diffusivity,
  for all simulations in Tab. \ref{tab:simu}.
  The scaling Eq. (\ref{eq:vrms}) is extremely well verified across two decades of thermal diffusivity,
  with a slight drop in the case of S0 and its 3D counterpart.
  The latter might be attributed to box-size effects that prevent the MTI from fully developing,
  as previously shown by \citet{mccourt13} using local models.
  The present study shows that such effects could be astrophysically relevant.
  The validity of Eq. (\ref{eq:vrms}) suggests that the MTI
  could drive turbulent motions of about a few hundreds of kilometers per second
  in the outermost regions of galaxy clusters,
  which correspond to Mach numbers $\mathcal{M}\sim 0.3$,
  as illustrated on the top right donut in Fig. \ref{fig:MTIfields}.
  Despite their potential strength, such MTI motions would be hardly detectable at the edges of galaxy clusters,
  on account on the ICM low emissivity there.
  The latter drops as $\propto\rho^2$: beyond the beige dashed line at $0.6R_\vir$ on the top right donut,
  the future ATHENA/X-IFU observatory would collect less than $1\times10^6$ photons in the $0.2-12$ keV range
  during a 1-Ms observation \citep{kempf23},
  given the X-IFU specifications before the recent design-to-cost exercise \citep{barret23}.
  In Section \ref{sec:pturb}, the associated induced level of non-thermal pressure support
  is further discussed, along with the possible caveats of this determination with our modelling approach.

  The same exercise is then performed, on the bottom panel in Fig. \ref{fig:scalings},
  to assess the dependency of the APE (i.e. thermodynamic fluctuations)
  on the volume-averaged thermal diffusivity.
  Similarly to the case of the turbulent velocity, the APE seems to follow a power-law
  $E_A \propto \chi$ over two decades in thermal conductivity.
  This scaling slightly differs from that proposed by \citet{pl22b} ($E_A\propto\chi^{3/4}$).
  Our scaling is used in Section \ref{sec:convection} to predict the 
  level of convective flux that should be expected from MTI-driven turbulence.

  Finally, directly comparing the global averages of the 3D simulations to their 2D counterparts
  (red and blue markers at the same $\volave{\chi}$ in Fig. \ref{fig:scalings}) reveals a systematic bias,
  which is particularly visible in the case of the APE (which is lower in 3D than in 2D).
  Similarly to the APE, the kinetic energy of the 3D runs is always below than that of their 2D counterparts,
  while their typical injection length is usually slightly above that of the 2D runs.
  We think that this effect is purely geometric:
  the geometry and the dimension of the magnetic field vector
  must somehow differ between 2D and 3D simulations.
  This could for example explain the lesser efficiency of the Braginskii heat flux to enforce
  the isothermality required by the MTI to drive buoyancy (see the discussion in Section \ref{sec:dmlt}).
  Mathematically speaking, the augmented dimensionality simply induces different prefactors
  in the relation $\varepsilon_i\propto-\varepsilon_\kappa\propto-\mathcal{A}$
  (through their dependence on the unit magnetic field vector $\hatb$),
  which may explain the systematic biases observed in Fig. \ref{fig:scalings} between the 3D and 2D simulations.

\subsection{Boundary conditions, thermodynamic profiles, thermal wind, and energy flux budget at saturation}
\label{sec:energyflux}
  \citet{parrish05,parrish07} first highlighted that the MTI has the apparent ability
  to bring the initial temperature gradient close to isothermality,
  especially when adiabatic boundaries are imposed
  (homogeneous Neumann BC on the temperature).
  In the design of our simulations, we thus tested several types of BC while
  paying special attention to the evolution of the background thermodynamic profiles with time.
  Accordingly, we found that imposing Neumann (zero-flux) BC on the temperature field leads to 
  almost flat temperature distributions at the end of the simulations with the highest conductivities,
  or partial and still ongoing flattening in the case of the runs with the lowest conductivities.
  We argue however that this feature is only an indirect consequence of the MTI;
  the latter only creates a background diffusive heat flux by radially re-orienting the magnetic field lines,
  and does not marginalise the profile by its own convective heat transport.
  We note that isotropic turbulence would have the same effect, expect perhaps less efficiently so.
  Indeed, we found that the timescale of this isothermalisation process
  is simply governed by the thermal diffusion time.
  For instance, we performed similar runs in which we replaced Braginskii by isotropic heat flux.
  Such simulations exhibit a similar flattening of the temperature profile,
  and most importantly, with very similar timescales,
  while no turbulence is occurring.
  This behaviour was first seen in simulations by \citet{parrish08},
  but only later elucidated by \citet{parrish12b}.
  Therefore, the flattening of the ICM temperature profile should not be seen
  as a specific property of the MTI saturation mechanism,
  since any isotropically tangled magnetic fields
  (which are most likely found in astrophysical galaxy clusters,
  contrary to purely isothermal horizontal magnetic field lines)
  of any turbulent origin would induce similar behaviours,
  to within a geometric suppression factor.
  As pointed out by \citet{parrish12b},
  the genuine problem of this kind of simulations lies in the fact that they do not self-consistently
  resolve the physics of both the MTI and the accretion shock,
  which however acts as the main driver of the temperature gradients in the ICM \citep{mccourt13}.
  Imposing inhomogeneous Dirichlet BC on the temperature then provides a convenient and consistent way
  to provide our numerical ICM atmosphere access to a large energy reservoir
  that has been filled, all along the cluster dynamical hierarchical accretion history
  (which is not self-consistently modelled here),
  thanks to heating by the accretion shock.
  This is the path we took in the present work.

\begin{figure}
\centering
\includegraphics[width=0.95\hsize]{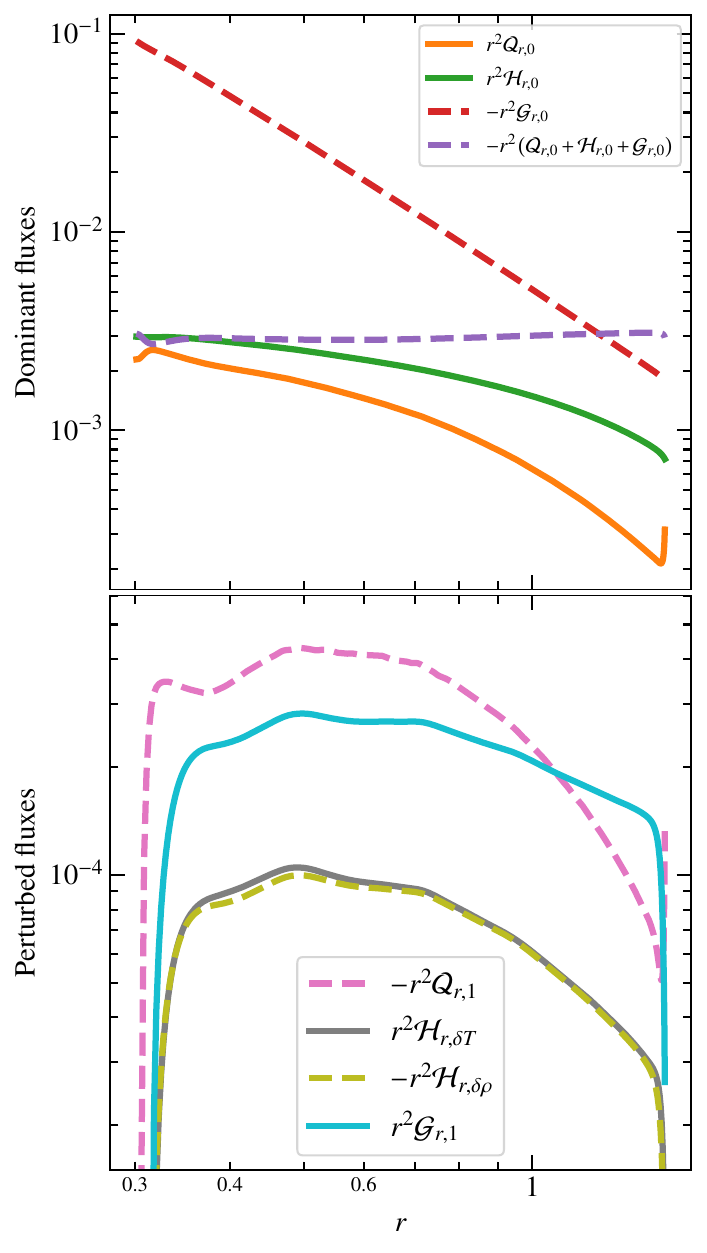}
\caption{
  Top: dominant equilibrium between Braginskii heat, enthalpy and gravitational
  energy fluxes, Eqs. (\ref{eq:qr0})-(\ref{eq:hr0})-(\ref{eq:gr0}),
  in the energy flux budget Eq. (\ref{eq:radfluxes}) at quasi steady-state, for the 3D simulation F0.
  Bottom: subdominant perturbed energy fluxes, Eqs. (\ref{eq:qr1})-(\ref{eq:hr1})-(\ref{eq:gr1}),
  for the same run.
  Dashed lines are indicative of negative quantities whose sign has been switched
  for the purpose of visualisation.
}
\label{fig:fluxes}
\end{figure}

  Nevertheless, even with such a BC, non-trivial saturation
  of the thermodynamic profiles still occurs
  in our simulations.
  In the top panel of Fig. \ref{fig:fluxes}, we show the leading radial fluxes
  transporting energy through the solution region of F0
  after a quasi-stationary regime of MTI saturation is reached.
  The presence of a strong advective flux $\mathcal{H}_{r,0}$, Eq. (\ref{eq:hr0}), can be somewhat surprising,
  but is actually easily explained.
  Our initial HSE state is not at thermal equilibrium in presence of any heat flux in the radial direction,
  but since the initial magnetic field lines are purely horizontal and the radial heat flux is shut off, it is nevertheless initially at rest.
  At the end of the linear phase, the opening of the magnetic field lines by the MTI in the radial direction induces 
  a Braginskii heat flux in this direction.
  However, such a non-zero radial heat flux destabilises the initial profile by bringing it out of HSE
  and causes a subsequent uniform radial bulk motion, which we refer to as a thermal wind.
  Such a dynamics is quite common in planetary atmospheric physics:
  upward thermal conduction of energy deposited deep into the atmosphere can drive a wind that tends
  to escape the planet's gravitational field \citep{johnson10}.
  This model of atmospheric escape is usually referred to as the slow hydrodynamic escape \citep{volkov11},
  and it is straightforward to explain, at least qualitatively.
  If a shell of stratified fluid heated from below is receiving more energy
  than it can thermally diffuse to cool down,
  it tends to expand. In doing so however, the pressure stratification induces anisotropy
  and prevents the shell from expanding towards the inner region.
  Such an expansion takes then the form of an outward wind, escaping the gravitational potential.
  In our simulations, the relative strength of the thermal wind
  (with respect to the MTI turbulence) is quantified through the ratio
  $\volave{\vv_r}/\volave{\vv_{r,\rms}}$ in Tab. \ref{tab:2Dresult}.
  The latter ratio increases with the thermal diffusivity because $\volave{\vv_r}\propto\chi$
  (this result is empirical but physically in line with our interpretation of the wind origin),
  while for the MTI, $\volave{\vv_{r,\rms}}\propto\chi^{1/2}$.

  This bulk advective heat flux
  was already seen in the simulations of \citet{parrish08},
  and deemed to be astrophysically irrelevant in that same work
  on the basis that the dark matter profile should somehow be able
  to adjust to the change in the ICM structure so as to limit this fluid advection.
  We do agree with the fact that this thermal wind might only bear a limited astrophysical relevance,
  but not with the physical mechanism invoked to justify this:
  the dynamics of the dark matter could hardly be set by that of the baryons,
  while the inverse is certainly true on account to their respective mass budgets
  \citep[e.g. gas sloshing in galaxy clusters;][]{johnson12,zuhone13}.
  Consequently and similarly to the possible flattening of the temperature profile,
  we rather see this wind as a consequence
  of not resolving,
  in isolated cluster simulations,
  the whole gravitational structure formation flows,
  whose interplay with other physical mechanisms
  would otherwise self-consistently set the thermodynamic ICM gradients
  without necessarily resorting to a thermal wind.
  We do not rule out however possible short episodes of thermal wind in the ICM, for example
  when AGN is feeding back more energy in the cluster core than the ICM plasma
  is able to thermally diffuse,
  although Bremsstrahlung cooling might also be at play in this process.
  Additionally, we found the structure of this wind to be critical
  for the shape of the thermodynamic profiles at saturation,
  in the runs with the highest conductivities $\kappaeff$
  (see the green thermodynamic profiles in Fig. \ref{fig:profiles} for the 2D run S0),
  but not that much for runs with moderate to low thermal diffusivities
  (yellow curves on the same plot for the 3D run F0).
  The wind structure is also very dependent on the BC imposed on the stratified thermodynamic quantities
  (either inhomogeneous Dirichlet or a local HSE BC, as described by \citeauthor{zingale02} \citeyear{zingale02}),
  and on the velocity field (either homogeneous Dirichlet or Neumann),
  which may be indicative of the limited relevance of this effect from an astrophysical viewpoint.
  We provide an additional study of the influence of those BCs on the wind,
  and on the MTI-driven turbulence itself more generally, in Appendix \ref{app:zingale}.

  In addition to heat, 
  the mass transfer induced by this thermal wind advects gravitational energy.
  The ordering $E_K,E_M \ll E_I, \left|E_G\right|$ found in the ICM,
  impose the predominance of these three fluxes (heat, enthalpy and gravitational) over the others
  (kinetic, viscous, Poynting, see Appendix \ref{app:energetics}).
  Interestingly, the top panel in Fig. \ref{fig:fluxes} shows that, in our 3D simulation F0,
  the evolution of the background thermodynamic profiles saturates by reaching
  a non-trivial dominant balance between the leading fluxes (purple dashed line).
  The same goes for the 2D run S0 (not shown).
  In the latter case, the new equilibrium profile (green curves in Fig. \ref{fig:profiles})
  looks like an intermediate state
  between the initial atmosphere (blue curves) and a hypothetical ICM atmosphere that would be both
  at hydrostatic and thermal equilibria (red curves)
  in presence of a radial background heat flux with Spitzer conductivity.
  Accordingly, the outermost shell of fluid exhibits a decreasing entropy profile,
  which violates Schwarzschild's criterion for adiabatic convective stability.
  However, this fluid shell is very thin, and subject to both anisotropic heat flux and viscosity.
  It is therefore very unclear whether such a region would actually be unstable to standard thermal convection,
  in addition to the MTI, or not,
  since negative entropy gradient is only a necessary condition
  for the onset of standard thermal convection.

  In the fiducial 3D run F0, a subdominant balance between the perturbed fluxes
  induced by MTI-driven turbulence is partially achieved.
  This balance is visible in the bottom panel in Fig. \ref{fig:fluxes}.
  The negative perturbed conductive flux $\mathcal{Q}_{r,1}$, Eq. (\ref{eq:qr1}),
  and the convective gravitational flux $\mathcal{G}_{r,1}$, Eq. (\ref{eq:gr1}), partially compensate each other,
  while the perturbed convective heat fluxes $\mathcal{H}_{r,\delta T}$
  and $\mathcal{H}_{r,\delta \rho}$ cancel each other at each radii,
  on account of the weakly compressible nature of the flow.
  The negative sign of $\mathcal{Q}_{r,1}$ translates the fact that, in MTI-driven turbulence,
  the temperature fluctuations tends to fight the background temperature gradient,
  by rearranging according to Eqs. (\ref{eq:mlT})-(\ref{eq:satMTI1}).
  We note however, that the convective energy fluxes, $\mathcal{H}_{r,\delta T}$, $\mathcal{H}_{r,\delta\rho}$,
  and $\mathcal{G}_{r,1}$ are small with respect to the background conductive heat flux $\mathcal{Q}_{r,0}$.
  These results reinforce our argument that the isothermalisation of the temperature profile
  is not the dominant, specific, process saturating the MTI;
  and that the latter is rather the marginalisation of the total Braginskii flux,
  as diagnosed through the APE budget in Sect. \ref{sec:fidevolution}.
  In Section \ref{sec:convection} below,
  we provide an explanation for these weak levels of convective energy fluxes,
  and discuss the implication of this result for galaxy clusters.

\section{Implications for ICM energetics and dynamics}
\label{sec:discussion}
  In this section,
  we show how the two different descriptions of the MTI at saturation, presented in Section \ref{sec:satMTI},
  can be reconciled into a single diffusive mixing-length theory (DMLT).
  We then make use of this theory
  to predict the levels of
  convective fluxes and turbulent pressure support produced by the MTI in the ICM.

\subsection{A diffusive mixing-length theory of the MTI}
\label{sec:dmlt}
  Here, we attempt to unify both MTI saturation descriptions that have been previously reviewed in Section \ref{sec:satMTI}.
  A striking feature
  is the close resemblance
  of Eqs. (\ref{eq:mlT})-(\ref{eq:satMTI1}),
  especially if we assimilate $\nabla$ to $\ell_i^{-1}$,
  the injection length of turbulence in the theory of \citet{pl22a}.
  Taking this path strongly suggests that,
  for the MLT to correctly describe the MTI at saturation,
  the mixing length $\ell_m$ should be chosen as
  the injection length $\ell_i$.
  When doing so, Eq. (\ref{eq:mlv}) is necessarily equivalent to Eq. (\ref{eq:vrms}),
  since $\omega_T$ is taken to be the natural time scale of MTI-driven turbulence in the theory
  of \citet{pl22a,pl22b}.
  Whether $\omega_T$ and $\ell_i$
  (and not the same quantities multiplied by some opposite powers of $N/\omega_T$ for instance)
  are exactly the right quantities to be regarded as
  the time scale and the mixing length of such turbulence cannot be decided within the current work,
  because no parameter study varying independently $\omega_T$ and $N$ was carried out.
  However, this caveat is only a theoretical limitation,
  which has very limited to no impact on any astrophysical conclusions
  since those time scales are very similar (and thus $N/\omega_T\sim 1$) in galaxy clusters
  and for our initial HSE.

  Actually, another and perhaps better physically motivated reason why
  $\ell_i$, to within any reasonable power of $N/\omega_T$, should be taken as the mixing length $\ell_m$
  goes as follows.
  Using a Taylor expansion to write Eq. (\ref{eq:mlT}), we implicitly assumed that the perturbation was
  perfectly isothermal. In this framework, the mixing length $\ell_m$ should therefore be understood
  as the typical length over which any perturbation can stay isothermal
  with respect to the temperature of its initial magnetic field line,
  in the cartoon picture of the MTI at least.
  Given some typical time scale $\omega^{-1}$ of MTI-driven turbulence
  (whose precise value does not matter for the current discussion,
  it should be quite close to $\omega_T$ and $N$ in the ICM anyway),
  the typical length
  over which a fluid element stays thermally connected to the radius at which it started moving,
  during the time scale $\omega^{-1}$ of the motion,
  must be proportional to the conduction length, defined as $\ell_\chi = \left(\chi/\omega\right)^{1/2}$,
  to within possible geometrical prefactors.
  Such a choice for the mixing length $\ell_m$ embeds
  the diffusive nature of the MTI in a MLT, which is otherwise diffusion free
  and would therefore misrepresent the basic physics of the instability.
  We refer to this approach as a diffusive mixing-length theory (DMLT),
  which, satisfactorily, leads to similar dependencies of the MTI injection length
  and velocity fluctuation strength on the thermal diffusivity
  (by comparing Eq. (\ref{eq:mlv}) to Eq. (\ref{eq:vrms}) when $\ell_m$ $\propto$ $\ell_\chi$).

  Similarly, we can compare the scaling of the temperature fluctuation intensity
  derived from both the DMLT and the work of \citet{pl22a,pl22b}.
  In the case of the DMLT, we find $\left(\delta T/T_0\right)^2 \propto \chi$
  rather than $\left(\delta T/T_0\right)^2 \propto \chi^{3/4}$,
  the former scaling being also seemingly preferred in our simulations
  (bottom panel in Fig. \ref{fig:scalings}).
  This might be so far the only example of divergent result between our work and \citet{pl22a,pl22b},
  in which only the gradient of the APE is constrained thanks to phenomenological arguments:
  in their theory, the scaling of the APE is therefore numerically measured
  and not theoretically derived.
  Similarly to the issue of the precise MTI-driven time scale $\omega^{-1}$ and mixing length $\ell_m$,
  this discrepancy with our work is only of secondary importance regarding astrophysical conclusions.
  We therefore propose the three following scalings:
\begin{align}
\label{eq:dml}
  &\ell_i \approx 4.6 \left(\frac{\chi}{\omega}\right)^{0.5}, \\
\label{eq:dmlv}
  &\vv^2 \approx 0.13 \omega \chi, \\
\label{eq:dmlT}
  &\left(g_0\frac{\delta T}{T_0}\right)^2 \approx 0.03 \omega^3 \chi,
\end{align}
  where $\omega$ is some typical frequency of MTI turbulent motions
  (taken equal to $1 \ \mathrm{Gyr^{-1}}$ here),
  which is likely to be a reasonable mix of $\omega_T$ and $N$.
  In the next sections, we apply these equations derived from the DMLT of MTI saturation
  to assess both the amount of convective fluxes
  and non-thermal turbulent pressure support
  that can be expected from the MTI in the outskirts of galaxy clusters,
  thereby revisiting the estimates of \citet{parrish12b} and \citet{mccourt13}.

\subsection{Convective fluxes carried by the MTI}
\label{sec:convection}

  Before assessing the convective efficiency of the MTI,
  a specific issue first needs to be tackled.
  When previous authors referred to the convective flux induced by the MTI,
  they systematically meant the quantity
  named $\mathcal{H}_{r,\delta T} \propto \sphave{\delta T \delta\vv_r}$ in the current work,
  which is the expression of the convective flux usually found in stellar convection
  \citep{hurlburt84,cattaneo91,porter00}.
  But in all generality, two other convective fluxes arise from the strong cross-correlation
  between temperature, density and radial velocity fluctuations,
  namely $\mathcal{H}_{r,\delta\rho}\propto\sphave{\delta\rho\delta\vv_r}$
  and $\mathcal{G}_{r,1}$, Eq. (\ref{eq:gr1}),
  and those must also be accounted for in any energy flux budget.
  The latter two fluxes are obviously proportional to each other,
  but they are related through another and more subtle relation too,
  and only cancel each other for adiabatic stratification.
  Indeed, the combination of the first and second principles of thermodynamics for an atmosphere at HSE 
  can be rewritten \citep{hurlburt84,rieutord15}:
\begin{equation}
  \frac{\mathcal{R}\gamma}{\gamma-1} \frac{\dd\sphave{T}}{\dd r} + \frac{\dd\Phi}{\dd r} = \sphave{T} \frac{\dd\sphave{S}}{\dd r},
\label{eq:hse}
\end{equation}
  and therefore,
\begin{equation}
  \mathcal{H}_{r,\delta\rho} + \mathcal{G}_{r,1} = \left[\left(\frac{\mathcal{R}\gamma}{\gamma-1} \sphave{T}^* + \Phi^*\right) + \int_{r^*}^r \sphave{T} \frac{\dd\sphave{S}}{\dd r}\right] \sphave{\delta\rho\delta\vv_r},
\label{eq:isen}
\end{equation}
  where $r^*$ is some reference radius
  (and $\sphave{T}^*$, $\Phi^*$ are the background temperature and gravitational potential at this radius).
  A natural choice for the latter radius is that at which the constant term
  $\left(\frac{\mathcal{R}\gamma}{\gamma-1} \sphave{T}^* + \Phi^*\right)$ cancels
  (which is always possible to find, because the gravitational potential is only defined up to a constant).
  From Eq. (\ref{eq:isen}), it is clear that $\mathcal{H}_{r,\delta\rho} \approx - \mathcal{G}_{r,1}$
  in the limit of an isentropic background stratification
  (which is the natural endpoint of saturation in standard thermal convection such as encountered in stellar convection).
  In galaxy clusters however, the background entropy profile is not adiabatic
  because it is imposed, at first order, by the dynamical hierarchical accretion history,
  and not by standard thermal convection.
  Then, there is no simple relation between $\mathcal{H}_{r,\delta\rho}$ and $\mathcal{G}_{r,1}$.
  Nevertheless, in the limit of mildly compressible MTI flows, we still have
  $\mathcal{H}_{r,\delta T} \approx -\mathcal{H}_{r,\delta\rho}$
  since pressure fluctuations are small
  (see for example the bottom panel in Fig. \ref{fig:fluxes}).
  The appropriate turbulent energy transport term to consider when referring to the convective flux
  induced by mildly compressible MTI turbulence in this context is therefore $\mathcal{G}_{r,1}$ rather than $\mathcal{H}_{r,\delta T}$.
  We now assess its strength with respect to the background heat flux.

  By essence, the DMLT introduced in Section \ref{sec:dmlt}
  to describe the large-scale saturation of the MTI
  is well-suited to calculate the induced level of convective flux.
  Indeed, convection-driven turbulence (to which the MLT usually applies) exhibits
  a significant cross-correlation
  between temperature and velocity fluctuations,
  and so does the MTI for which we numerically measured
  $\alpha\left(\delta\vv_r,\delta T\right)\approx 0.5$ (see Tab. \ref{tab:2Dresult}),
  on account of its buoyant nature.
  This exercise was previously performed by \citet{mccourt13},
  but with an incomplete understanding of the convective fluxes and MLT associated to MTI-driven turbulence.
  Using Eqs. (\ref{eq:dmlv})-(\ref{eq:dmlT}),
  we rederive the ratio of the convective flux of gravitational energy,
  Eq. (\ref{eq:gr1}),
  to the background Braginskii conductive heat flux,
  Eq. (\ref{eq:qr0}),
  as:
\begin{equation}
\begin{aligned}
  \frac{\mathcal{G}_{r,1}}{\mathcal{Q}_{r,0}}
  &= 7\% \left(\frac{\alpha\left(\delta\vv_r,\delta T\right)}{0.5}\right)
          \left(\frac{b_r^2}{0.5}\right)^{-1}
          \left(\frac{\delta\vv^2_r/\delta\vv^2}{0.5}\right)^{1/2}
          \left(\frac{n_0/\nel}{0.5}\right)
          \left(\frac{\omega^2}{\omega^2_T}\right) \\
  &\times \left(\frac{\kB T}{2 \ \mathrm{keV}}\right)^{-1}
          \left(\frac{\mu\mH\Phi}{2 \ \mathrm{keV}}\right).
\end{aligned}
\label{eq:ratioconvcond}
\end{equation}
  This result needs to be considered with care.
  Indeed, Eq. (\ref{eq:ratioconvcond}) only suggests that 
  the convective gravitational flux induced by the MTI in the ICM
  is very weak with respect to the background heat flux, $\mathcal{Q}_{r,0}$,
  contrary to previous claims regarding the convective heat flux \citep{parrish05,parrish07,mccourt11}.
  However, MTI-driven turbulence is known
  to significantly open the magnetic field lines in the radial direction.
  This radially biased magnetic field offers a stronger channel
  for the background heat flux to flow within the ICM periphery,
  in comparison to an isotropically tangled magnetic field,
  and must to some extent be attributed to the MTI itself.
  This ordering between the convective gravitational and conductive heat fluxes is specific to the MTI:
  classical convection is usually more efficient at transporting energy than conduction.
  In the Sun for instance, the convective envelope is unstable to thermal convection
  precisely because the effective conduction of heat\footnote{In the Sun, an effective thermal conduction is due to the radiative diffusion.}
  is too low to efficiently diffuse all the luminosity coming from the radiative zone,
  and other non-linear modes of energy transfer like convection naturally arise.
  We stress that this is definitely not the situation of the MTI in the ICM.
  Unlike standard thermal convection, the MTI needs fast parallel heat conduction to develop in the first place
  and to thrive at saturation, as shown by \citet{pl22a,pl22b}.
  With this in mind, Eq. (\ref{eq:ratioconvcond}) is only a natural consequence of the MTI diffusive nature
  and should thus not be surprising.
  
\subsection{Non-thermal pressure support generated by the MTI}
\label{sec:pturb}
  We now turn our attention to the non-thermal pressure support induced by the MTI in the outermost ICM.
  The scaling Eq. (\ref{eq:dmlv}),
  equivalently described by the DMLT or by the balance mechanism between the APE leading injection and dissipation rates \citep{pl22a,pl22b},
  offers a proxy to assess the ratio of non-thermal to thermal pressures,
  given thermodynamic profiles, for typical MTI flows in the ICM.
  Using typical values for the outermost regions $r\sim 1.5 R_\vir$ leads to:
\begin{equation}
\begin{aligned}
  \frac{p_\mathrm{turb}}{p_0} &=
  14\% \left(\frac{p_0}{1 \times 10^{-14} \ \mathrm{g/cm/s^2}}\right)^{-1} 
  \left(\frac{\omega}{1 \ \mathrm{Gyr^{-1}}}\right) \\
  &\times \left(\frac{\kappa}{1 \times 10^{11} \ \mathrm{g / cm/s^3/K}}\right)
  \left(\frac{n_0/\nel}{0.5}\right).
\end{aligned}
\label{eq:pturb}
\end{equation}
  Interestingly, this numerical value is non-negligible
  and is an indicator of moderate Mach numbers possibly induced by the MTI close to the virial shock.
  Nevertheless, we caution that this ratio is an increasing function of the radius.
  For example, it quickly drops to $5\%$ for ICM values taken at $r\sim R_\vir$,
  and it keeps doing so toward the innermost and denser regions of ICM haloes.
  The radial dependence of Eq. (\ref{eq:pturb}) should therefore carefully be accounted for
  before drawing definitive conclusions about the ability of the MTI to significantly bias
  the mass inferred from X-ray and SZ observations combined with the HSE assumption;
  especially since the highest values of this ratio are found in the least dense ICM regions,
  which weigh little in the mass budget of galaxy clusters.
  Another caveat to this result is the weakly to almost non-collisional nature of the ICM plasma
  at such large radii.
  For instance, well-known kinetic instabilities \citep{schekochihin05,schekochihin10,drake21}
  could efficiently suppress the thermal conductivity in weakly collisional plasmas,
  and therefore the strength of MTI-induced turbulence according to Eqs. (\ref{eq:dmlv})-(\ref{eq:dmlT}).
  Their interplay with the MTI is briefly discussed in Section \ref{sec:caveat}.

\section{Summary and conclusions}
\label{sec:conclusion}

  In this paper, we have studied turbulence induced,
  in a global spherical stratified ICM atmosphere,
  by the saturation of the magneto-thermal instability,
  a magneto-buoyant instability likely to impact the internal dynamics and energetics
  of galaxy cluster outskirts.
  This work specifically aimed at
  bridging the gap between
  idealised simulations of MTI-driven turbulence in local and global ICM models
  \citep{parrish07,parrish08,mccourt11,parrish12b,pl22a,pl22b},
  and large-scale simulations of structure formation through gravitational collapse
  including anisotropic thermal conductivity \citep{ruszkowski11}.
  The main results are threefold:
\begin{itemize}
\item The MTI saturation mechanism
      introduced by \citet{pl22a,pl22b} under the Boussinesq approximation
      is found to result from a balance between the APE leading injection and dissipation rates.
      This balance still holds for strong stable entropy stratification
      and collisional regimes of unsuppressed parallel thermal conduction.
      A positive APE injection rate is only possible because the MTI saturates
      by marginalising the anisotropic Braginskii heat flux.
      We emphasise that this saturation mechanism does not imply
      the marginalisation of the background temperature gradient itself by the MTI,
      which was previously suggested to be the natural endpoint of this instability \citep{parrish05,parrish07}.
\item The mixing-length theory (MLT),
      proposed by \citet{parrish12b,mccourt13} to describe the large-scale dynamics of the MTI,
      and which seemed unrelated to the previous local saturation mechanism of \citet{pl22a,pl22b},
      can be made consistent with it
      by choosing the mixing length to be proportional to the conduction length $\ell_\chi \propto \sqrt{\chi}$.
      This is a way to re-introduce the diffusive nature of this instability in a MLT,
      which remains otherwise essentially diffusion-free
      and cannot thus account for the physics of the MTI.
      We refer to this approach as a diffusive MLT (DMLT).
      However, since $\ell_\chi \sim H_T \sim H_p$ in the ICM anyway for a complete Spitzer conductivity,
      choosing $\ell_m \sim H_p$ over $\sim \ell_\chi$ does not significantly impact the numerical predictions.
      This is the reason why the previous estimate of the convective over conductive energy fluxes ratio
      of few percents by \citet{mccourt13} is well in line with ours, Eq. (\ref{eq:ratioconvcond});
      though we showed that the right convective flux to consider is $\mathcal{G}_{r,1}$, Eq. (\ref{eq:gr1}),
      rather than $\mathcal{H}_{r,\delta T}$.
\item Finally, we confirmed a clear dependency of the turbulent injection length
      and strength on the thermal conductivity (over more than two decades of conductivity),
      that can be inferred from the saturation mechanism of \citet{pl22a,pl22b}.
      This dependency is well incorporated by our DMLT too.
      For instance, in the 2D simulation with a complete Spitzer conductivity,
      the MTI drives cluster-size motions at Mach numbers $\mathcal{M}\sim0.3$.
\end{itemize}

\subsection{Implications for the ICM}
\label{sec:implication}

  We made use of these results, and especially of the scaling Eq. (\ref{eq:dmlv}),
  to deduce the levels of convective flux and non-thermal turbulent pressure support
  associated with MTI-driven turbulence in the ICM periphery.
  We found, both theoretically and numerically,
  that, in standard B-MHD with no suppression of conduction,
  the MTI can sustain mildly compressible turbulence with a typical length scale of the size of the cluster,
  with Mach numbers as high as $\mathcal{M}\sim 0.3$
  in the outermost regions.
  In this case,
  the ratio of the non-thermal to thermal pressures $p_\mathrm{turb}/p_0 = \gamma \mathcal{M}^2$
  can reach values up to $15\%$ in regions close to the virial shock,
  see Eq. (\ref{eq:pturb}) for a precise estimate.
  We stress however that this ratio is a quickly increasing function of the radius,
  mostly because the background pressure $p_0$ is exponentially decreasing.
  The radial variation of the MTI dynamical properties was discussed by \citet{kempf23},
  who assessed the possible detectability of this instability in a Perseus-like cluster at $r\sim 0.25 R_\vir$
  with the future X-ray spectrometer X-IFU \citep{barret23} onboard ATHENA.
  In \citet{kempf23}, we adopted a local radius-by-radius approach to the MTI dynamical properties.
  The present global study shows however that, in case of a realistic Spitzer conductivity,
  the MTI can drive buoyant plumes extending over the full ICM outskirts.
  This could possibly slightly boost the estimates of the velocity signal and of the injection length
  expected from the innermost ICM regions, that are at the same time unstable to the MTI
  and available for X-ray observation with the ATHENA/X-IFU
  (inside the area delimited by the beige dashed line on the top right donut in Fig. \ref{fig:MTIfields}).
  This is why, directly applying Eq. (\ref{eq:pturb}) locally at a given radius
  may not be justified with the highest values of thermal conductivity,
  as far as the assessment of the turbulent pressure support is concerned.

  Another application of Eq. (\ref{eq:vrms}) is for the assessment of the convective energy transport
  induced by the MTI.
  We showed that the ratio between the convective gravitational
  and the background Braginskii heat fluxes hardly exceeds $7\%$,
  Eq. (\ref{eq:ratioconvcond}),
  which is pretty low for a physical process which is often referred to as convection,
  or as a convective instability.
  Whether the latter denomination is justified depends on the meaning we give it:
  does it refer to the ability of the induced-turbulence to carry significant convective flux
  (and if so, are we speaking about transport of heat or gravitational energy, or both?),
  or to the efficient coupling between temperature and radial velocity fluctuations?
  In standard thermal convection like in the Sun, both are true,
  and the energy carried by the convective flux is a mix between the gravitational and thermal components.
  However, because of the diffusive nature of the MTI, which needs fast parallel conduction to trigger turbulence,
  MTI-driven convection is not as efficient as conduction at transporting heat across the ICM.
  Referring to the MTI as a convective instability might therefore be misleading in this respect,
  and emphasising the buoyant nature of the instability instead seems more appropriate.

\subsection{Caveats and future perspectives for ICM studies}
\label{sec:caveat}
  We now bring to light other possible applications of the MTI in an astrophysical context,
  along with the associated caveats.
  First, and most importantly,
  similar further studies about the interplay between the MTI and other forms of turbulence driven by different processes
  are needed to draw definitive conclusions about the actual impact of the MTI
  on the dynamics and physics of galaxy clusters.
  Second, there is also a strong need to understand the interaction of this fluid instability
  with possible harmful effects of kinetic micro-instabilities occurring in high-beta plasmas,
  that we completely overlooked in this paper,
  in order to assess whether this instability is relevant at all in ICM outskirts.
  This is a research area under intense investigation.
  For example, \citet{berlok21} demonstrated the robustness of the MTI against mirror modes
  suppressing the heat conductivity,
  while \citet[][]{perrone24a,perrone24b} recently shed light on
  the possibly deadly non-linear feedback of whistler modes on the MTI,
  which strongly depends on the ability of the latter to sustain sufficient levels of magnetic field.

  Speaking of which, a promising area of application
  for MTI studies of the kind conducted here
  lies into the possible dynamo effect that the MTI may drive.
  Whether the MTI can,
  through a small-scale fluctuation dynamo process,
  bring the very weak primordial magnetic fields
  up to equipartition with the kinetic energy, as observed in close clusters,
  is an outstanding question
  which we plan to further investigate in future work.
  Here, a thorough study of realistic magnetic field amplification by MTI-driven turbulence
  was prevented because of the $\Pm=1$ regime chosen,
  and the subsequent moderate magnetic Reynolds numbers $\Rm = \mathcal{O}(100)$,
  of the intermediate resolutions of the few 3D runs presented.
  We further note that the choice of isotropic over Braginskii viscosity,
  must certainly modify the turbulent dynamics at the viscous scale,
  on which fluctuation dynamo critically depends \citep{rincon19},
  and should therefore be carefully studied in future dynamo works.

\subsection{Possible astrophysical applications of the APE}
  In this paper, the concept of available potential energy (APE) has been introduced as a useful tool to understand
  the physics and saturation of the MTI in the ICM.
  While a complete and accurate description of this quantity in galaxy clusters is far beyond the scope
  of the current study (further details are still given in Appendix \ref{app:ape}),
  we argue that the APE is of primary importance, not only for the ICM internal dynamics,
  but also for the observation of this dynamics in X-ray.
  Indeed, pending the future generation of X-ray spectrometers
  like Resolve \citep{ishisaki18} or the X-IFU \citep{barret23},
  indirect measurements of the ICM kinematics are only achievable
  through the observation of X-ray surface brightness fluctuations
  \citep{zhuravleva15, dupourque23}.
  In the ICM, density perturbations with respect to the background profile
  are usually assumed to be directly related
  to the velocity fluctuations through their respective spectra \citep{zhuravleva14}.
  In this methodology,
  the interplay between kinetic and available potential energies is clearly at stake,
  and is actually taken to be scale-independent.
  This assumption has been checked to some extent in hydrodynamic simulations of forced stratified turbulence
  with conditions representative of the ICM \citep{gaspari14,mohapatra21}.
  It remains however to be tested in presence of magneto-buoyant instabilities
  like the MTI or the HBI, since we demonstrated that, in MTI-induced turbulence,
  kinetic energy is driven by, rather than driving, density fluctuations (i.e. APE).

  We argue that this framework could also be applied
  to stellar, solar and planetary atmospheric dynamics to further understand the energy transport
  through, possibly double-diffusive, convective fluxes within those astrophysical objects,
  thanks to the fully compressible equations of fluid dynamics.
  Like in the case of the MTI,
  the saturation of double-diffusive instabilities might be similarly controlled by the injection rate of APE
  (more details are given in Appendix \ref{app:ape}).
  In the same manner,
  the energetics of the $\kappa$-mechanism, which is responsible for the pulsation of Cepheid stars
  \citep{eddington17},
  that are used as standard candles to build the cosmic distance ladder \citep{riess19},
  seems to be ruled by the conversion of BPE into APE
  \citep[though not formally interpreted as such;][]{gastine08}.
  Besides, \citet{tailleux24a} recently proposed static energy asymptotics as a method to better
  constrain the energetics of the different classes of Boussinesq and anelastic approximations that are used
  in the ocean and atmospheric modelling community.
  We argue that similar models used in astrophysics,
  like the adiabatic anelastic approximation used in stellar physics \citep{brun04,miesch05},
  in protoneutron star dynamics \citep{reboul-salze22},
  or in planetary interiors \citep{gastine12,gastine12a,gastine21},
  could benefit from the same treatment to improve our understanding of their energetics,
  building back on the early insights from \citet{braginsky95}.

\begin{acknowledgements}
  The authors thank the anonymous referee for valuable comments
  which helped to improve the clarity of this paper.
  The work presented here benefited from fruitful exchanges with
  Nicolas Clerc,
  Jean-Baptiste Durrive,
  Thomas Guillet,
  Christopher Howland,
  Laurène Jouve,
  Henrik Latter,
  Geoffroy Lesur,
  Leo Middleton,
  Yves Morel,
  Lorenzo Perrone,
  Michel Rieutord,
  and
  Remi Tailleux.
  The authors are grateful to R. Tailleux for his careful reading of Appendix B,
  and for bringing the relevant notion of extrinsic state function to their attention.
  J.M.K. has been partially supported through the grant EUR TESS N°ANR-18-EURE-0018
  in the framework of the Programme des Investissements d'Avenir.
  This work was granted access to the HPC resources of CALMIP under the allocation 2023/2024-P16006,
  and to those of CINES under the allocations 2024-A0160415090 made by GENCI.
  We are grateful for the work of the associated support teams.

  J.M.K: Ce travail est dédié à la mémoire de
  mon défunt professeur Jean-Pierre Simond,
  dont le sens physique a beaucoup apporté à ma formation scientifique.
\end{acknowledgements}

%
%

\bibliographystyle{aa}
\bibliography{ref}

\begin{appendix} 
\section{Energy flux budget in B-MHD}
\label{app:energetics}

  The gravitational work on the right-hand-side of Eq. (\ref{eq:Etotcons})
  is related to the gravitational energy $E_G=\rho\Phi$ according to:
\begin{equation}
  \frac{\partial E_G}{\partial t} + \vnabla \cdot \left(E_G\,\vvv\right) = \rho \frac{\partial\Phi}{\partial t} + \rho\,\vvv\cdot\vnabla\Phi,
\label{eq:Eg}
\end{equation}
  where the first term on the right-hand-side is zero in our case,
  since the NFW gravitational potential of dark matter is independent of time in our models.
  A total budget equation of radial energy fluxes also including the gravitational energy can thus be obtained
  by spherically averaging the sum of Eqs. (\ref{eq:Eg}) and (\ref{eq:Etotcons}):
\begin{align}
  &\frac{\partial}{\partial t} \sphave{E+E_G} \\
  &+ \frac{1}{r^2} \frac{\partial}{\partial r} \left( r^2 \left[
    \mathcal{Q}_r
  + \mathcal{H}_r
  + \mathcal{G}_r
  + \mathcal{K}_r
  + \mathcal{D}_r
  + \Pi_r
  \right] \right)
  = 0, \nonumber
\end{align}
where:
\begin{align}
  &\mathcal{Q}_r = \sphave{-\kappa\left(\hatb\cdot\vnabla T\right)b_r}, \\
  &\mathcal{H}_r = \frac{\mathcal{R}\gamma}{\gamma-1} \sphave{p \vv_r}, \\
  &\mathcal{G}_r = \Phi \sphave{\rho \vv_r}, \\
  &\mathcal{K}_r = \frac{1}{2} \sphave{\rho \vv^2 \vv_r}, \\
  &\mathcal{D}_r = \sphave{\vec{\Sigma}\cdot\vvv} \cdot \vec{e_r}, \\
  \label{eq:poynting}
  &\Pi_r = \frac{1}{\mu_0}\sphave{\vec{\mathcal{E}}\times\vec{B}} \cdot \vec{e_r},
\end{align}
  are the spherically-averaged
  radial heat, enthalpy, gravitational, kinetic, viscous and Poynting fluxes, respectively.
  In Eq. (\ref{eq:poynting}), $\vec{\mathcal{E}} = - \vvv \times \vec{B} + \eta \vec{j}$
  is the local electric field of the plasma.
  In the ICM, the kinetic and magnetic energy volume densities are comparable,
  and several orders of magnitude smaller than the internal and gravitational energy densities.
  We therefore expect the kinetic, viscous and Poynting fluxes to contribute very little to
  the total energy flux budget;
  so we discard them from the current discussion.
  Focusing on the remaining heat, enthalpy, and gravitational fluxes,
  we note that they are themselves a sum of very different contributions,
  that can easily be highlighted when all physical quantities $X$
  (other than the unit vector $\hatb$) are further decomposed
  into their spherically-averaged
  and fluctuating\footnote{By definition, $\sphave{\delta X}$ is zero.} parts $\sphave{X}+\delta X$.
  More specifically, the radial heat flux $\mathcal{Q}_r$ can be rewritten as
  the sum of background $\mathcal{Q}_{r,0}$ and perturbed $\mathcal{Q}_{r,1}$ heat fluxes:
\begin{equation}
  \mathcal{Q}_r = -\sphave{\kappa b_r^2}\partial_r\sphave{T} - \sphave{\kappa\bigl(\hatb\cdot\vnabla\delta T\bigr)b_r}.
\end{equation}
  The radial enthalpy flux $\mathcal{H}_r$ is the sum of an advective $\mathcal{H}_{r,0}$
  and two convective $\mathcal{H}_{r,1}=\mathcal{H}_{r,\delta T} + \mathcal{H}_{r,\delta\rho}$ heat fluxes:
\begin{equation}
  \mathcal{H}_r = \frac{\mathcal{R}\gamma}{\gamma-1}\Bigl(\sphave{\rho T}\sphave{\vv_r}
                +                                   \sphave{\rho}\sphave{\delta T \delta\vv_r}
                +                                   \sphave{T}\sphave{\delta \rho \delta\vv_r}\Bigr),
\end{equation}
  and similarly for the gravitational flux,
  which is the sum of an advective $\mathcal{G}_{r,0}$ and a convective $\mathcal{G}_{r,1}$ fluxes:
\begin{equation}
  \mathcal{G}_r = \Phi \Bigl(\sphave{\rho} \sphave{\vv_r}
                +             \sphave{\delta\rho\delta\vv_r}\Bigr).
\end{equation}

\section{Introduction to the concept of available potential energy}
\label{app:ape}

  In this work, we introduced the notion of available potential energy (APE)
  to understand the non-linear saturation of the MTI.
  The goal of this section is to better describe this physical quantity,
  since it might look, at first glance, unfamiliar in astrophysics.

  Given the close relation, Eq. (\ref{eq:hse}), between the specific internal energy (or enthalpy)
  and specific gravitational energy for an atmosphere at HSE in a gravitational field $\Phi$,
  \citet{lorenz55} proposed to process both internal and gravitational energies
  as a single type of energy in a stratified atmosphere.
  The net gain of kinetic energy through adiabatic buoyancy motions comes at the expense of both energies.
  Therefore, \citet{lorenz55} defined the sum of the gravitational potential and internal energies
  as the total potential energy (TPE) $E_P = E_G + E_I$.
  However, by definition of a stably stratified atmosphere at HSE (with respect to thermal convection),
  it seems that none of the gravitational energy $E_G$ of such an equilibrium state
  can be converted into kinetic energy, since small velocity perturbations will never grow
  but rather be systematically damped.
  This illustrative example shows that, in stratified fluid dynamics,
  not all the gravitational energy (and \textit{a fortiori} not all the TPE)
  is available for conversion into kinetic energy.
  This motivated \citet{lorenz55} to introduce the notion of available potential energy (APE) $E_A$,
  which quantifies the part of the TPE that is available for reversible conversions into kinetic energy.
  Formally, the APE is the difference between the TPE
  and the background potential energy (BPE) $E_B$,
  which is defined as the minimum TPE that can be obtained from an adiabatic (and thus reversible)
  redistribution of masses at HSE in the initial gravitational field $\Phi$ \citep{winters95}.
  With such a definition, the APE is nothing else than the amount of buoyant, and adiabatic compressive, work
  that must be exerted against the background stratification to bring a fluid element
  from its current physical condition (i.e. its height and pressure)
  to the position and pressure it would have in the reference state,
  that realises the BPE \citep[that is the minimum TPE;][]{tailleux18}.
  Guided by such considerations, \citet{andrews81} first derived local evolution equations
  for both the APE and BPE, absent any non-ideal effect.
  \citet{tailleux18} later generalised these equations for a viscous and thermally conducting fluid.
  Here, we extend it again, to include Ohmic dissipation:
\begin{align}
\label{eq:ea}
  &\frac{\partial E_A}{\partial t}
  + \vnabla \cdot \left(E_A\,\vvv + \delta p\,\vvv + \frac{\delta T}{T}\vec{q}\right) \\
  &= \delta\rho\,\vvv\cdot\vnabla\Phi + \vvv\cdot\vnabla\delta p
  + \vec{q}\cdot\vnabla\left(\frac{\delta T}{T}\right)
  + \frac{\delta T}{T}\left(\vec{\Sigma}:\vnabla\vvv + \eta\mu_0 j^2\right) + \mathcal{F}, \nonumber \\
\label{eq:eb}
  &\frac{\partial E_B}{\partial t}
  + \vnabla \cdot \left(E_B\,\vvv + p_*\,\vvv + \frac{T_*}{T}\vec{q}\right) \\
  &= -\vec{q}\cdot\vnabla\left(\frac{\delta T}{T}\right)
  + \frac{T_*}{T}\left(\vec{\Sigma}:\vnabla\vvv + \eta\mu_0 j^2\right) - \mathcal{F}, \nonumber
\end{align}
  with $\vec{q}$ some heat flux,
  and where the fluctuating quantities here are defined with respect to the reference state $\cdot_*$
  that realises the BPE.
  In Section \ref{sec:diag} however, these fluctuations are defined locally,
  with respect to the spherically-averaged quantities $\sphave{\cdot}$.
  Therefore, we implicitly chose the background profiles as the reference state,
  which is however not guaranteed to be that of minimum TPE in this case.
  What exact information (and if any, which amount) is lost when choosing an arbitrary reference state,
  rather than the sorted state, in such studies is still to be elucidated.
  In the same manner, it must be noted that the quadratic APE introduced in Eq. (\ref{eq:quadape}) 
  is only an approximation valid in the limit of small displacements and density perturbations \citep{roullet09, tailleux13},
  and does certainly not hold for arbitrarily large displacements \citep{holliday81,andrews81},
  despite having been used as such in simulations of compressible stratified ICM turbulence here or in \citet{mohapatra20}.
  The APE and the BPE are both extrinsic state functions, that is a joint function depending on
  the state variables describing the local fluid properties,
  as well as the properties of the environment within which it is embedded.
  Therefore, an additional exchange term $\mathcal{F}$ between APE and BPE
  accounting for the temporal variation of the environment itself, that is the fluid as a whole,
  appears in both Eqs. (\ref{eq:ea})-(\ref{eq:eb}) \citep{tailleux18}.
  Though we acknowledge that a complete volume-averaged budget of APE would require taking into account the diverse boundary fluxes
  and the non-local exchange term too, we just used the latter equation to illustrate the basic phenomenology of the MTI energetics
  in Section \ref{sec:fidevolution}.

  From Eqs. (\ref{eq:ea})-(\ref{eq:eb}), it is readily shown that the sum of the evolution equations for
  the available $E_A$ and background potential energy $E_B$ gives the right evolution equation for the
  total potential energy $E_P$,
  under the assumption that the reference state is at HSE 
  ($\rho_*\vnabla\Phi + \vnabla p_*=0$):
\begin{equation}
  \frac{\partial E_P}{\partial t}
  + \vnabla \cdot \left(E_P\,\vvv + p\,\vvv + \vec{q}\right)
  = \rho\,\vvv\cdot\vnabla\Phi + \vvv\cdot\vnabla p
  + \vec{\Sigma}:\vnabla\vvv + \eta\mu_0 j^2.
\label{eq:ep}
\end{equation}
 The conversion rate $\varepsilon_A$, defined as:
\begin{equation}
  \varepsilon_A = \vec{q}\cdot\vnabla\left(\frac{\delta T}{T}\right),
\end{equation}
  is absent of the latter evolution equation.
  This quantity is then a conversion rate between APE and BPE, due to heat flux.
  On the contrary, the last two terms on the right-hand side of Eq. (\ref{eq:ea}) are only indirectly related
  to such a conversion process: they occur in the conversion rate of the sum of the kinetic energy and the APE into BPE.
  In the ocean and in the atmosphere, the conversion rate between APE and BPE $\varepsilon_A$
  is usually referred to as a dissipation rate (going from APE into BPE) because,
  most of the time, it is negative in physical conditions representative of these systems,
  and even negative-definite in case of an incompressible Boussinesq stratified single-component fluid \citep{winters95}.
  However, \citet{middleton20,middleton21,tailleux24} recently showed that this rate
  can turn positive when double-diffusive instabilities are developing.
  In this work, we clearly demonstrated that this conversion rate can also be positive (from BPE to APE)
  in the case of the MTI, and that it even controls its saturation.
  This feature should not be surprising given the obvious similarities that the MTI shares
  with other double-diffusive instabilities.
  For example, their maximum growth rates do not depend on the level of diffusivity itself,
  and they all modify the Schwarzschild's criterion for stability,
  on account on their diffusive nature (although different diffusive processes are at stake in each case).
  We therefore speculate that the energetics of double-diffusive instabilities \citep{garaud18},
  like the fingering instability for instance \citep{traxler11,fraser24}, may be similarly controlled
  by the positive injection rate of BPE into APE.
  Finally, we stress that the possible indirect conversion of BPE into KE (through APE)
  is only made possible here thanks to non-ideal effects;
  while \citet{lorenz55} first defined the APE so as to distinguish the part of the TPE
  that is available for reversible conversion into kinetic energy from the part that is not (the BPE).
  In thermodynamics, available energy is sometimes referred to as exergy,
  defined as the maximum useful work that can be produced by a physical system as it is brought to equilibrium
  with its environment through an ideal process \citep{cengel11}.
  When the physical system under consideration is a perfect stratified fluid,
  the APE seems to play exactly this role \citep{kucharski97}.
  As highlighted all along this paper however,
  the relaxation of the MTI is very far from being adiabatic.
  Even more surprising, in this case,
  kinetic energy (i.e. useful work) can only be produced thanks to non-ideal effects
  such as parallel thermal conduction.
  Therefore, new ways of defining the exergy budget of a thermally conducting dilute stratified fluid,
  like the ICM, should be explored to circumvent this theoretical limitation.
  For instance, partitioning the TPE through an isothermal, rather than adiabatic,
  rearrangement of masses might represent a promising avenue in this direction.

\section{Influence of the boundary conditions applied on the stratified thermodynamic fields}
\label{app:zingale}

\begin{figure*}[h]
\centering
\includegraphics[width=\hsize]{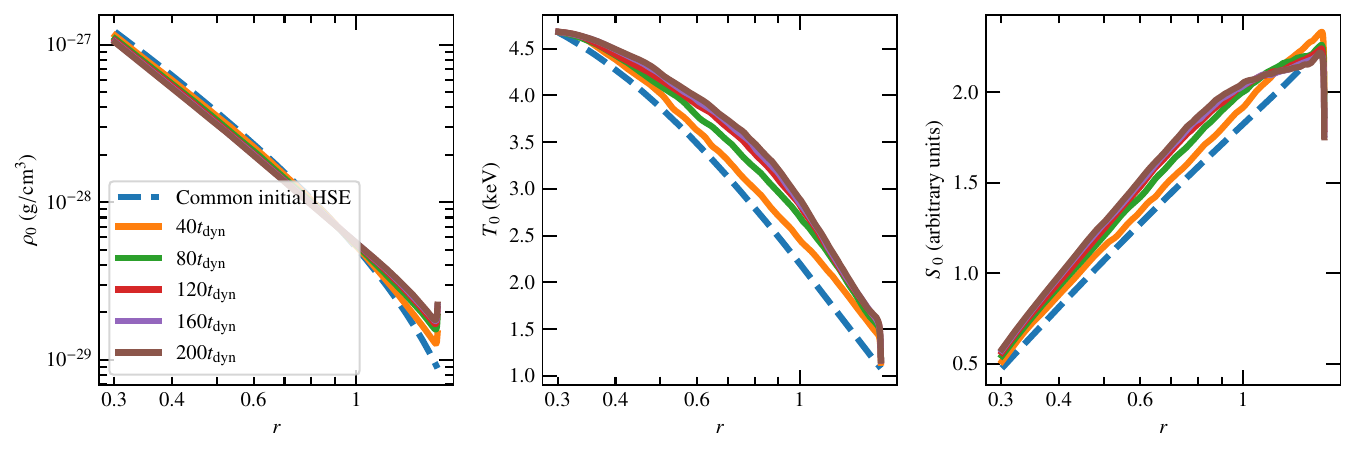}
\caption{
  In dashed lines: initial stratified ICM atmosphere;
  density $\rho_0$ (in $\mathrm{g/cm^3}$, left), temperature $T_0$ (in $\mathrm{keV}$, center) and
  entropy $S_0$ (right) profiles.
  In plain lines: same profiles but displayed at various time during the simulation
  with the local HSE BC of \citet{zingale02}.
  This figure should be compared to Fig. \ref{fig:profiles}, which was obtained with the BCs
  described in Section \ref{sec:numeric}.
}
\label{fig:profileszingale}
\end{figure*}

\begin{figure}[h]
\includegraphics[width=0.82\hsize]{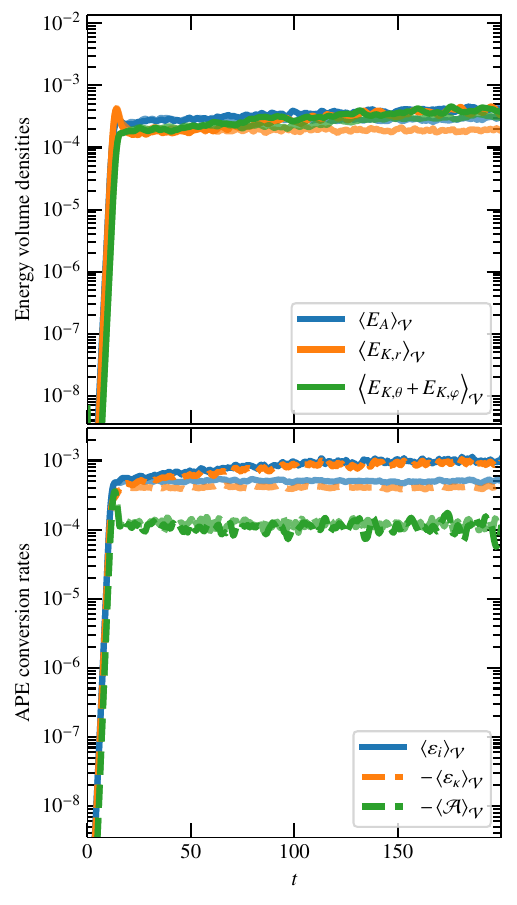}
\caption{
  Top: volume-averaged time evolution
  of the available potential energy density $\volave{E_A}$,
  and of the different contributions to the
  kinetic $\volave{E_K}$ volume energy density in the run with the local HSE BC of \citet{zingale02}.
  Bottom: volume-averaged time evolution of the
  APE injection rate $\volave{\varepsilon_i}$,
  APE thermal dissipation rate $-\volave{\varepsilon_\kappa}$,
  and the (opposite amount of) reversible buoyancy work $-\volave{\mathcal{A}}$
  during the same run.
  Dashed lines are indicative of negative quantities whose sign has been switched
  for the purpose of visualisation.
  Opaque curves are for the current simulation,
  while transparent curves are for the 2dDf1e-2 run, which was obtained with the BCs
  described in Section \ref{sec:numeric}.
}
\label{fig:timezingale}
\end{figure}

  In all the simulations presented in Tab. \ref{tab:simu},
  we set indirect inhomogeneous Dirichlet BC on the temperature field
  by imposing, all along the runs, both the density and the pressure in the ghost cells
  to be those of the initial ICM atmosphere at HSE.
  The main limitation of such BCs is the possible development of small boundary layers 
  on the density and temperature fields, as can be seen at the inner and outer boundaries
  on the background density profile, $\sphave{\rho}$, in the 2D run S0
  (green curve on the leftmost panel in Fig. \ref{fig:profiles}),
  or on the background heat flux, $\mathcal{Q}_{r,0}$, in the 3D run F0
  (yellow curve on the top panel in Fig. \ref{fig:fluxes}).

  As an attempt to dispose of such boundary layers, we implemented a first-order local HSE BC,
  as suggested by \citet[][Eq. 41]{zingale02}:
\begin{equation}
  p_{i+1} - p_i = - \left(\Phi_{i+1} - \Phi_i\right) \frac{\rho_{i+1} + \rho_i}{2},
\end{equation}
  which must be solved for the two unknowns $p_{i+1}$ and $\rho_{i+1}$
  in the case of the outer (rightmost) ghost cells,
  or for both $p_i$ and $\rho_i$ at the inner boundary.
  Imposing inhomogeneous Dirichlet BC on the temperature field in the ghost cells
  is then enough to determine both quantities, thanks to the equation of state.
  With these new BCs on the stratified thermodynamic fields, 
  coupled to homogeneous Dirichlet BC on the radial velocity,
  we reran the 2D simulation 2dDf1e-2 for 200 dynamical times $\tdyn$.
  In Fig. \ref{fig:profileszingale}, we plot the background density, temperature, and entropy profiles
  at various times during the simulation;
  while in Fig. \ref{fig:timezingale}, we display the time evolution of the diverse energetic diagnostics
  introduced in Section \ref{sec:diag}.
  Both figures should be compared to their counterparts,
  Figs. \ref{fig:profiles} and \ref{fig:timeF0}, respectively.
  Several features of this new run are noteworthy.
  First, these new BCs do not prevent the formation of boundary layers
  on the density, temperature, and entropy profiles.
  Second, the thermodynamic profiles seem to keep evolving all along the simulations and to never saturate,
  or at least to do so on a timescale much larger than the usual simulation,
  like the 2dDf1e-2 or F0 runs.
  As a result of the constantly evolving temperature profile, MTI-driven turbulence never reaches
  a proper quasi-steady state during the simulation.
  As seen on the middle panel in Fig. \ref{fig:profileszingale}, the average temperature through the domain
  is globally increasing with time.
  Indeed, a local HSE BC tends to suppress the thermal wind,
  whose intensity is then much lower in this run than in its counterpart 2dDf1e-2.
  Consequently, the thermal wind is no longer able to advect enough internal energy
  out of the domain: the fluid is constantly heated, without being sufficiently cooled down,
  since the conductive heat flux is larger at the inner than at the outer boundary.
  This result suggests that the thermal wind acts a self-regulation thermal mechanism,
  in a fluid heated from below.
  Despite the highlighted boundary layers and the unsteady saturation,
  the regime of MTI-induced turbulence, and its subsequent phenomenology,
  remain qualitatively unchanged in volume,
  with very similar zeroth-order levels of saturation than in the initial run 2dDf1e-2,
  as seen from the comparison between the opage curves (current simulation),
  and their transparent counterparts (2dDf1e-2 run),
  in Fig. \ref{fig:timezingale}.
  Therefore, our conclusions on the nature of magneto-thermal turbulence
  are quite independent on the BCs used in this work.

\end{appendix}
\end{document}